%% file: paper.tex
\documentclass[usenatbib, twocolumn]{mn2e}
\usepackage{epsfig}
\usepackage{amsmath}
\usepackage{lscape}
\usepackage{longtable}
\usepackage{anysize}
\usepackage{color}
\usepackage{footnote}
\usepackage{ulem}
\usepackage{natbib}
\input{macro}

\begin{document}
%

\title[Star Formation and Assembly of Galaxies]
      {Star Formation and Stellar Mass Assembly in Dark Matter Halos: From Giants to Dwarfs}

\author[Lu et al.]
       {\parbox[t]{\textwidth}{
        Zhankui Lu$^{1}$\thanks{E-mail: lv@astro.umass.edu},
        H.J. Mo$^{1}$,
        Yu Lu$^{2}$,
        Neal Katz$^{1}$,
        Martin D. Weinberg$^{1}$,
        Frank C. van den Bosch$^{3}$,
        Xiaohu Yang$^{4,5}$}\\
           \vspace*{3pt} \\
$^1$Department of Astronomy, University of Massachusetts, Amherst MA 01003-9305, USA\\
$^2$Kavli Institute for Particle Astrophysics and Cosmology, Stanford, CA 94309, USA\\
$^3$Astronomy Department, Yale University, P.O. Box 208101, New Haven, CT 06520-8101, USA\\
$^4$Center for Astronomy and Astrophysics, Shanghai Jiao Tong University,
n     Shanghai 200240, China\\
$^5$Key Laboratory for Research in Galaxies and Cosmology, Shanghai Astronomical 
    Observatory, \\ \, Nandan Road 80, Shanghai 200030, China}


\date{}
\pagerange{\pageref{firstpage}--\pageref{lastpage}}
\pubyear{2014}

\maketitle 

\label{firstpage}


\begin{abstract}  
The empirical model of \citet{Lu14} is updated with recent data
of galaxy stellar mass functions (SMFs). The model predicts that
the slope of galaxy SMFs at $z>2$ should be quite steep at the low mass
end, beyond the current detection limit, and
it is a strong prediction that can be tested against future observations.
The model is used to investigate the galaxy star formation and
assembly or merger histories in detail. 
Most of the stars in cluster centrals, corresponding to BCGs
in observations, formed earlier than $z\approx 2$ but have been 
assembled much later. Typically, they have experienced $\approx5$
major mergers since their star formation was quenched.
Milky Way mass galaxies have had on-going star formation 
without significant mergers since $z\approx 2$, and are thus 
free of significant (classic) bulges produced by major mergers.
Dwarf galaxies in haloes with $M_{\rm h} < 10^{11}h^{-1}M_{\odot}$ 
or $M_{\star} < 10^{9}M_{\odot}$ have experienced a star formation burst
at $z > 2$, followed by a nearly constant star formation rate after $z = 1$,
and the stellar age decreases with stellar mass, contrary to 
the `downsizing' trend for more massive galaxies.
Major mergers are not uncommon during the early burst phase and 
may result in the formation of old spheroids in dwarf galaxies. 
We also characterise the stellar population of halo stars
in different halos.
\end{abstract}

\begin{keywords}
galaxies: haloes ---
galaxies: formation --- 
methods: statistical 
\end{keywords}


\section{Introduction}
\label{sec_intro}

Given that the hierarchical formation of dark matter haloes is the
backbone of galaxy formation, and that dark matter is the
dominant mass component in the Universe, understanding the origin and
diversity of the galaxy population requires a solid understanding of
the connection between galaxies and dark matter haloes. 
The last decade has seen tremendous progress in this area.
By combining the galaxy-dark matter connection as a function of redshift
with halo merger trees, which describe the hierarchical assembly of
dark matter haloes, one obtains a statistical description of how
galaxies assemble their stellar mass over time. Such an `empirical'
approach is intuitive and transparent in that it describes galaxy
evolution within the natural framework of hierarchical structure
formation.  This method was developed by \citet{Conroy07}, \citet{Conroy09} and
\citet{Yang09}, and has since been applied in numerous studies
\citep[e.g.][]{Moster10, Moster13, Behroozi10, Behroozi13a,
Behroozi13b, AvilaReese11, Yang12, Yang13, Wang13, Bethermin13}.
An important new insight that has resulted from this
`empirical modelling' approach is that there is an amazing amount of
`regularity' and `simplicity' in galaxy demographics. In particular,
the relatively mild evolution in the $\Mstar - \Mhalo$ relation with
redshift is found to translate into an instantaneous star formation
efficiency (i.e. the star formation rate divided by the baryon
accretion rate) that peaks at a characteristic halo mass of $\sim
10^{12} \Msunh$, over the entire epoch from $z = 8$ to the present
day, and almost all star formation has occurred inside haloes that lie
within the narrow range $10^{11} \Msunh \la \Mhalo \la 10^{12}
\Msunh$ \citep[][]{Yang12, Yang13, Leauthaud12a, Leauthaud12b,
Behroozi13a, Behroozi13b, Moster13}. In addition, these studies have
shown that virtually all stellar mass is assembled {\it in situ}:
merging is only a significant channel of mass assembly for the most
massive galaxies. In what follows we will refer to this picture of
galaxy formation as the `Slow-Evolution' model, to emphasise that it
suggests little evolution in the halo mass dependence of the
star formation efficiency.

A recent study by the authors \citep{Lu14} has argued for an important
revision to this `Slow-Evolution' picture. Using data on the faint-end of
the luminosity function of cluster galaxies, they show that
low mass haloes ($\Mhalo \la 10^{11} \Msunh$) have to form stars
efficiently, but only at high redshift ($z \ga 2$). In particular,
they identified $z \simeq 2$ as a new characteristic epoch in galaxy
formation, where there is a fairly sudden transition in the star
formation efficiency of low mass haloes. 
This transition leads to some interesting predictions, e.g.
a significant old stellar population in present-day dwarf galaxies 
with $M_{\star} \le 10^{8}\Msunhh$ and steep low end slopes at high redshift 
of the galaxy stellar mass and star formation rate functions.
Interestingly, recent work based on deeper high redshift surveys 
provide some evidence for such a steepening.

The goal of the paper is to update the model of \citet{Lu14}
with more recent observational data and 
to characterise in detail the star formation and merger histories 
of galaxies across cosmic time with the updated model. 
We address questions related to {\it in
 situ} star formation versus accretion, downsizing both in star
formation and assembly and mergers of galaxies and their roles in shaping
galaxy morphology, and the prevalence of halo stars.  
Wherever possible we will contrast the differences between 
the `Slow-Evolution' model and the modified picture 
advocated by \citet{Lu14}.

This paper is organised as follows.  Our method to model the star
formation-halo mass connection is described in \S\ref{sec_model}. The
overall trends in the star formation and assembly histories are
described in \S\ref{sec_general}.  In \S\ref{sec_histories}, we
describe the total star formation and assembly histories as a function
of halo (stellar) mass, paying particular attention to {\it in situ}
star formation versus accretion, the breakdown of downsizing, and the
global star formation rate density in connection to the ionisation
state of the Universe. In \S\ref{sec_mergers}, we focus on galaxy
merger rates and their implications for the morphological transformation of
galaxies. In \S\ref{sec_starpopulation} we address the stellar
populations of galaxies as a function of their stellar mass, 
paying particular attention to the properties of halo stars.
Finally, we summarise our results in \S\ref{sec_summary}.

Throughout the paper, we use a $\Lambda$CDM cosmology with
$\Omega_{\rm m,0}=0.273$, $\Omega_{\Lambda,0}=0.727$, $\Omega_{\rm
b,0}=0.0455$, $h=0.704$, $n=0.967$ and $\sigma_{8}=0.811$.  
This set of parameters is from the seven year WMAP observations \citep{Komatsu11}.
In addition, unless stated otherwise, we adopt the stellar population
synthesis model of \citet{Bruzual03} and a Chabrier (2003) IMF.

\section{An Empirical Model of Star Formation in Dark Matter Halos}
\label{sec_model}

\subsection{The Model}

In this section we provide a brief description of the model of 
\citet{Lu14}, which we adopt here to make model predictions, 
referring the reader to the original paper for details. 

The hierarchical assembly of individual dark matter haloes is modelled 
using halo merger trees generated with the 
Monte-Carlo model of \citet{Parkinson08}, which
is based on a modified treatment of the extended Press-Schechter
formalism calibrated using $N$-body simulations
\citep[see][]{Cole08}. As shown in \citet{Jiang14}, the merger trees
obtained with this method are as accurate as those obtained from
high-resolution $N$-body simulations.

We assume that the star formation rate (SFR) of a central galaxy 
in a halo at a given redshift $z$ is determined by the virial mass of 
the host halo, $M_{\rm h}(z)$, and the redshift $z$, and hence
\begin{equation}\label{eqn_general01}
{\rm SFR} = {\dot M}_\star \left[ M_{\rm h}, z \right] \,,
\end{equation}
where $M_{\rm h} = M_{\rm h}(z)$ is the instantaneous halo mass.
\citet{Lu14} adopted the following functional form for the mass and
redshift dependence:
\begin{equation}\label{eqn_general02}
    {\dot M}_\star = {\cal E} {f_{\rm b} M_{\rm h} \over\tau}
    (X+1)^{\alpha}  \left(\frac{X+\mathcal{R}}{X+1}\right)^{\beta} 
    \left(\frac{X}{X+\mathcal{R}}  \right)^{\gamma} 
    \,,
\end{equation}
where ${\cal E}$ is a free parameter that sets the overall efficiency;
$f_{\rm b} = \Omega_{\rm b,0}/\Omega_{\rm m,0}$ is the cosmic baryon
mass fraction; and $\tau = (1/10\,H_0) \, (1+z)^{-3/2}$ roughly
describes the dynamical timescale of haloes at redshift $z$.
We define the quantity $X$ as $X \equiv M_{\rm h} / M_{\rm c}$
where $M_{\rm c}$ is a characteristic mass; and $\mathcal{R}$ is a
positive number smaller than $1$.  Hence, the SFR depends on
halo mass through a piece-wise power law, with $\alpha$, $\beta$, and
$\gamma$ being the three power indices in the three mass
ranges separated by the two characteristic masses, $M_{\rm c}$ and
$\mathcal{R} M_{\rm c}$.

\citet{Lu14} considered three different models. In Model I they
assumed that all the model parameters were independent of redshift.
However, they showed that such a model was unable to match 
simultaneously the observed galaxy stellar mass functions (SMFs)
in the redshift range between $z=0$ and $z=4$.  This can be remedied
by allowing $\alpha$ to depend on redshift according to
\begin{equation}\label{eqn_alpha}
  \alpha = \alpha_{0} (1+z)^{\alpha'} \,, 
\end{equation}
with both $\alpha_0$ and $\alpha'$ being free parameters. This Model II is
able to fit the SMFs from $z=0$ to $z=4$, and is very similar to the 
`Slow Evolution' model discussed above; virtually all
stars form in dark matter haloes in a narrow band of halo mass:
$10^{11} \Msunh \la M_{\rm h} \la 10^{12} \Msunh$.  However, this
model is unable to match the steep, faint-end upturn in the cluster
galaxy luminosity function at $z=0$ \citep{Popesso06}. 
This upturn requires an additional modification, namely that the 
parameter $\gamma$, which controls the star formation efficiency 
in low mass haloes, depends on redshift according to
\begin{equation}\label{eqn_gamma}
  \gamma = 
  \begin{cases}
    \gamma_{\rm a}   \, & \text{if}\, z < z_{c} \\
    (\gamma_{\rm a}-\gamma_{\rm b})
    \left(\frac{z+1}{z_{c}+1}\right)^{\gamma'} + 
     \gamma_{\rm b}  \, & \text{otherwise}\,,
  \end{cases}
\end{equation}
so that it changes from $\gamma_{\rm b}$ at high-$z$ to $\gamma_{\rm
  a}$ at low-$z$, with a transition redshift $z_{\rm c}$. Performing a Bayesian 
inference using  both the stellar mass functions from $z=0$ to
$z=4$ and the cluster galaxy luminosity function as data constraints,
\citet{Lu14} obtained a
model, Model III, that fits all the data. This model  differs from the
`Slow Evolution' model (and thus Model II) in that it predicts efficient
star formation in low mass haloes ($\Mhalo \la 10^{11}\Msunh$) but
only for $z > z_c \simeq 2$. 
As we will see later, the value of $z_c$ is quite well constrained. 
Our tests using a simpler model with $z_c=0$, shows that 
the model given by equation (\ref{eqn_gamma}) is favoured
by a factor $K \approx e^{15}$ in terms of the Bayes ratio
[see equation (13) in \citet{Lu14} for the definition],   
demonstrating clearly that the data prefer a non-zero $z_c$.  
More complicated models are not preferred by the data.

The above SFR model is only valid for central galaxies, i.e., galaxies
that reside at the centres of their dark matter host haloes, where
they act as the recipients of any new gas that looses its binding
energy.  Motivated by the notion that galaxies quench their star
formation once they become satellite galaxies
\citep[e.g.,][]{Balogh00,vdBosch08,Wetzel12}, \citet{Lu14} model the
SFR of satellite galaxies using a simple $\tau$ model:
\begin{equation}\label{SFR_satellite}
 {\dot M}_{\star,{\rm sat}}(t) = {\dot M}_{\star}(t_{\rm a}) 
       \exp\left( -\frac{t-t_{\rm a}}{\tau_{\rm sat}} \right) \,,
\end{equation}
where $t_{\rm a}$ is the time when the satellite is accreted into its
host and ${\dot M}_{\star}(t_{\rm a})$ is the SFR of the satellite galaxy
at $t=t_{\rm a}$. The quantity $\tau_{\rm sat}$ is the 
`quenching' time scale for satellite galaxies, which is modelled as
\begin{equation}
 \tau_{\rm sat} = \tau_{\rm sat,0} \exp\left( -\frac{M_\star}{M_{\star,c}} \right) \,,
\end{equation}
where $\tau_{\rm sat,0}$ is the exponential decay time for a satellite
galaxy with a stellar mass of $M_{\star,c}$; both $\tau_{\rm sat,0}$
and $M_{\star,c}$ are treated as free parameters. 
As discussed in \citet{Lu14}, the choice of $\tau_{\rm sat}$ is motivated by
the fact that a decreasing quenching time scale with 
stellar mass seems to be required by the data 
\citep{Peng10,Wetzel13}, but there is not really a sound physical
justification for our adopted form. While the predicted colours
and current star formation rates of satellite galaxies depend 
crucially on the quenching model, the predictions of 
the SMFs of galaxies that we are interested in 
here are insensitive to the adopted quenching model.

Each satellite galaxy is assumed to merge with the 
central galaxy of its host halo in
a time given by the dynamical friction time scale. During the merger,
only a fraction $f_{\rm TS}$ of the stellar mass of the satellite is
added to the central galaxy; the remainder is considered
`stripped' and becomes part of the `halo stars' \citep{Monaco06, Conroy07}. 
The fraction $f_{\rm TS}$ is treated as a free parameter. 

For each galaxy, we track its star formation ${\rm SFR}\left(t\right)$
and mass assembly history $M_{\star}\left(t\right)$.
At a time step $t_{i}$, we obtain the stellar mass following
\begin{eqnarray}
 \label{eq:assembly}
 M_{\star}(t_{i}) 
 & = & \int_{0}^{t_{i}} {\rm SFR}(t) R\left(t_{i}-t\right) dt \nonumber\\
 &   & +\sum_a f_{\rm TS} M_{{\rm s},\star}(t_{i})\,,
\end{eqnarray}
which takes into account both the passive evolution and accretion
of satellite galaxies denoted by the subscript `s'.
The summation is over all satellites.
$R\left(t_{i}-t\right)$ is the remaining fraction of stellar mass
for a stellar population of age $t_{i}-t$, and is adopted from the 
stellar evolution model of \citet{Bruzual03} assuming a Chabrier IMF \citep{Chabrier03}.
The total star formation rate
is simply the sum of all the progenitors
\begin{equation}
 \label{eq:sfh}
 {\rm SFR}(t_i) = {\rm SFR}_{\rm c}(t_i) + \sum f_{\rm TS} {\rm SFR}_{\rm s}(t_i)\,,
\end{equation}
where the central is denoted by a subscript `c' and satellites by `s'.

The luminosity can be calculated in the same way as the stellar mass
by replacing $R(t_i-t)$ with the light to mass ratio of a simple stellar 
population. We use the metallicity - stellar mass relation from \citet{Gallazzi05}.
Note that the $z$- and $r$-band luminosities are quite insensitive to 
the assumed metallicity.

\subsection{Updating model parameters with recent observational data} 

\begin{table}
\caption{The constrained model parameters of Model II, in terms 
         of the means and the variances.
         The observational constraints used  are listed in the first row.
         B12 is for \citet{Baldry12}, S12 for \citet{Santini12}
         and T14 for \citet{Tomczak14}.
         $M_{\rm c}$ is in units of $10^{10}\Msunh$, 
         and $M_{\rm *,c}$ is in units of $10^{10}\Msunhh$.
        }
  \setlength\tabcolsep{5pt}
  \begin{tabular}{ccc}
   \hline
     & SMF ($z\approx 0$,B12) & SMF ($z\approx 0$, B12)\\
     & SMF ($z\ge 1$, S12)    & SMF ($z\ge 1$, T14)\\
   \hline
   \hline
   Parameter    & mean$\pm\sigma$  & mean$\pm\sigma$  \\
   \hline

   $\alpha_{0}$ & $-3.7 \pm 0.82$  & $-3.4 \pm 0.84$\\ [1ex]
 
   $\alpha'$    & $-0.46\pm 0.11$  & $-0.45\pm 0.11$\\ [1ex]

   $\beta$      & $3.4 \pm 0.86$   & $2.6 \pm 0.99$\\ [1ex]

   $\gamma$     & $0.89 \pm 0.63$  & $1.3\pm 0.77$\\ [1ex]

   $\log_{10}(M_{\rm c})$ & $1.7 \pm 0.13$ & $1.8 \pm 0.18$  \\ [1ex]

   $\log_{10}({\cal R})$  & $-1.1\pm 0.34$ & $-1.1\pm 0.45$  \\ [1ex]

   $\log_{10}({\cal E})$  & $0.30 \pm 0.27$& $0.017\pm 0.27$  \\ [1ex]

   $\log_{10}(H_{0}\tau_{\rm sat,0})$ & $-0.98 \pm 0.17$ & $-0.85\pm 0.13$ \\ [1ex]

   $\log_{10}(M_{\rm *,c})$          & $0.81 \pm 0.42$  & $0.84\pm 0.38$ \\ [1ex]

   $f_{\rm TS}$                      & $0.36 \pm 0.17$  & $0.38\pm 0.15$ \\ 
  \hline
  \hline
\end{tabular}
 \footnotetext[1]{foot}
\label{modelii}
\end{table}

\begin{table}
  \caption{The same as Table~\ref{modelii}, but for Model III,
           and the cluster galaxy luminosity function \citep{Popesso06}
           is also used as a constraint.}
  \setlength\tabcolsep{5pt}
  \begin{tabular}{ccc}
   \hline
   \hline
     & SMF ($z\approx 0$, B12)& SMF ($z\approx 0$,B12) \\
     & SMF ($z\ge 1$, S12)    & SMF ($z\ge 1$, T14)   \\
   \hline
   \hline
   Parameter    &  mean$\pm\sigma$    &  mean$\pm\sigma$    \\
   \hline

   $\alpha_{0}$ &  $-3.0 \pm 1.0$     & $-2.7 \pm 0.85$     \\ [1ex]
 
   $\alpha'$    &  $-0.36 \pm 0.16$   & $-0.37 \pm 0.10$    \\ [1ex]

   $\beta$      &  $3.7 \pm 0.73 $    & $3.9 \pm 0.69$      \\ [1ex]

   $\gamma_{a}$ & $2.0 \pm 0.55 $     & $0.58 \pm 0.39$     \\ [1ex]

   $\gamma_{b}$ & $-0.84 \pm 0.14$    & $-0.90 \pm 0.08 $   \\ [1ex]

   $\gamma'$    & $-4.4 \pm 0.52$     & $-4.2 \pm 0.62$     \\ [1ex]

   $z_{\rm c}$  &  $1.8 \pm 0.31$    & $2.0 \pm 0.38$      \\ [1ex]  

   $\log_{10}(M_{\rm c})$ & $1.6 \pm 0.15$    & $1.6 \pm 0.13$    \\ [1ex]

   $\log_{10}({\cal R})$  & $-0.86 \pm 0.18$  & $-0.88 \pm 0.17$  \\ [1ex]

   $\log_{10}({\cal E})$  & $0.20 \pm 0.29$   & $0.07 \pm 0.27$      \\ [1ex]

   $\log_{10}(H_{0}\tau_{\rm sat,0})$ & $-0.90 \pm 0.16$  & $-0.74 \pm 0.04$  \\ [1ex]

   $\log_{10}(M_{\rm *,c})$ & $0.34 \pm 0.28$  & $0.36 \pm 0.17$    \\ [1ex]

   $f_{\rm TS}$             & $0.44 \pm 0.22$  & $0.34 \pm 0.19 $   \\ [1ex]
 
   $\log_{10}(e_{\rm M})$   & $0.15 \pm 0.04$  & $0.16 \pm 0.03 $   \\
  \hline
  \hline
\end{tabular}
\label{modeliii}
\end{table}

\begin{figure*}
 \centering
 \includegraphics[width=0.9\linewidth]{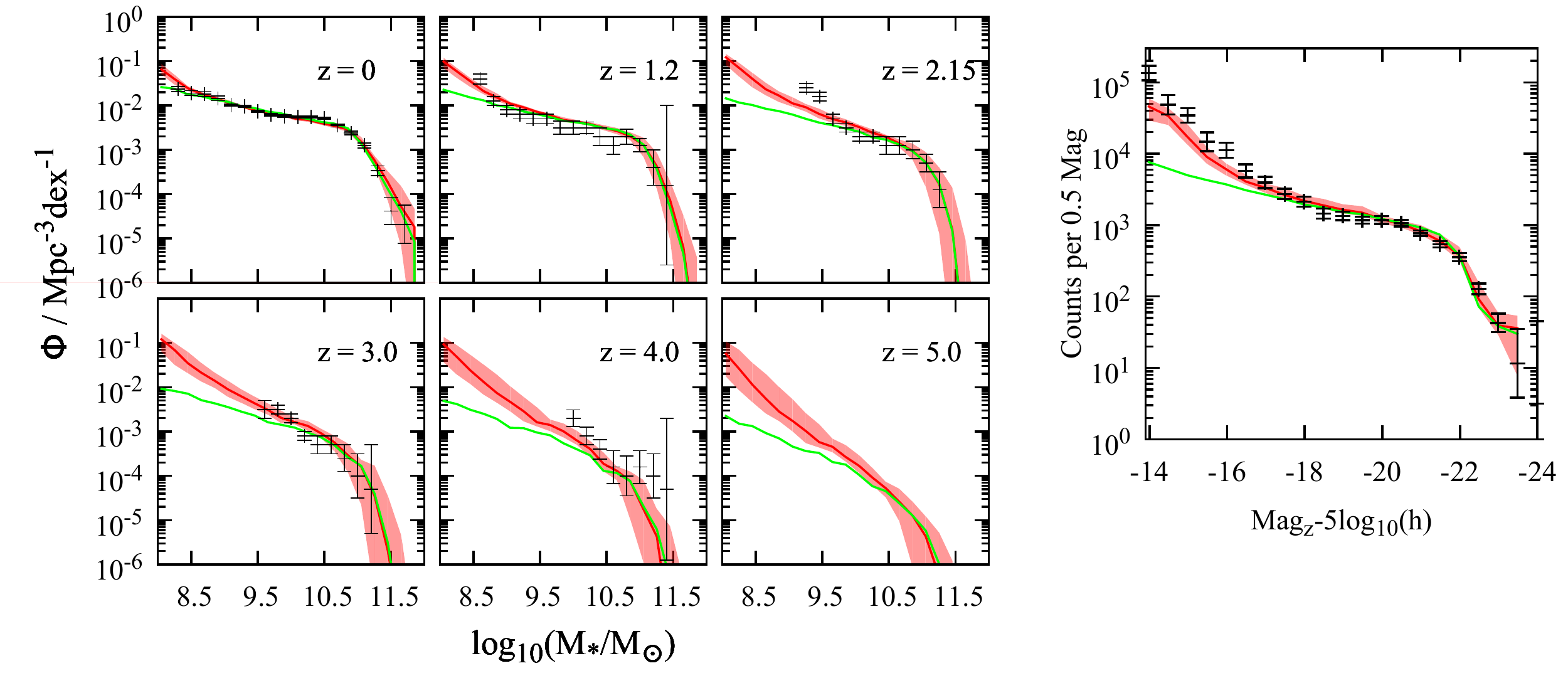}
 \caption{Comparison between the posterior predictions and
          the constraining data (points with error bars). 
          The high-$z$ SMFs are from \citet{Santini12}.
          The red curves (best fit) and the bands (95\% credible
          intervals) are predictions of Model III, while the green
          curves are the predicted means of Model II.}
 \label{ppc01}
\end{figure*}
\begin{figure*}
 \centering
 \includegraphics[width=0.9\linewidth]{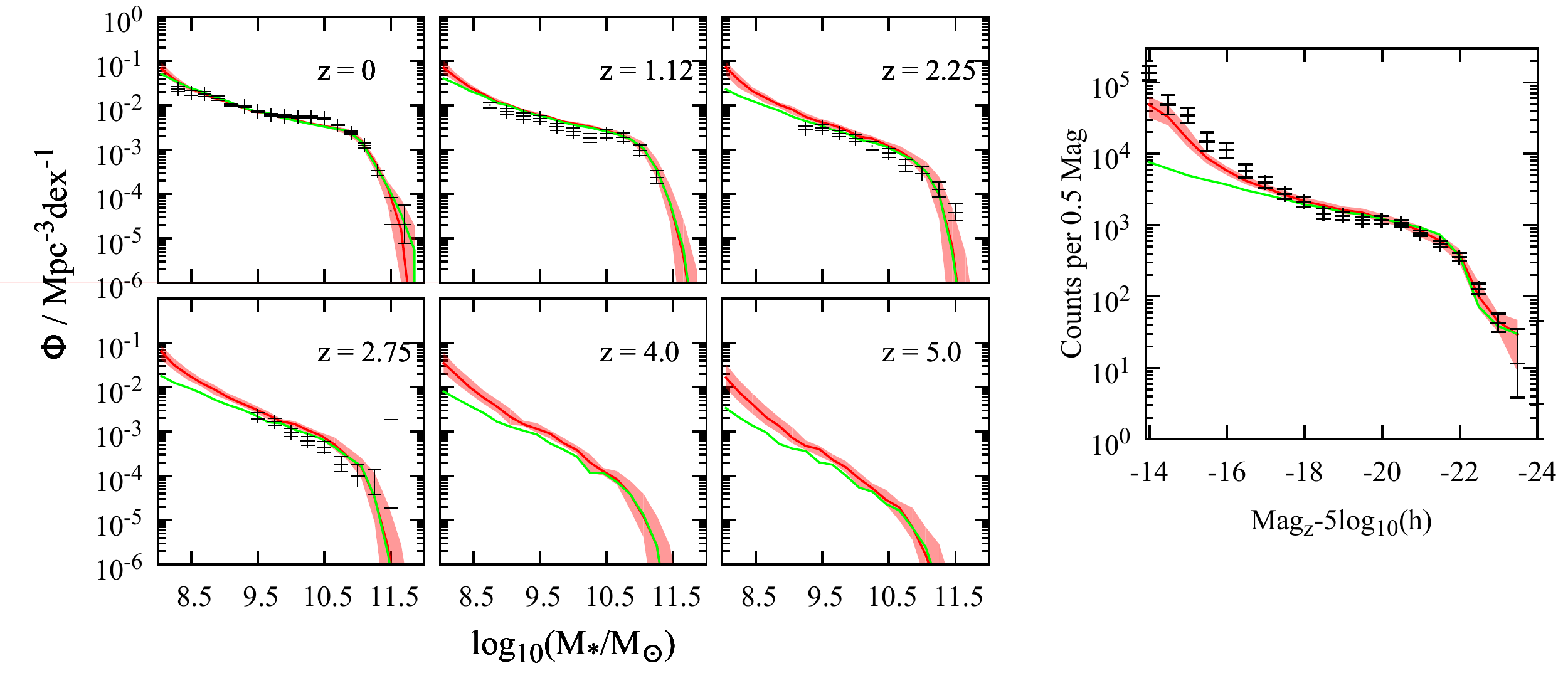}
 \caption{The same as Fig.~\ref{ppc01} but the high-$z$ constraining
          data are from \citet{Tomczak14}.}
 \label{ppc02}
\end{figure*}

\begin{figure*}
\centering
 \begin{minipage}{0.48\linewidth}
 \includegraphics[width=1.0\linewidth]{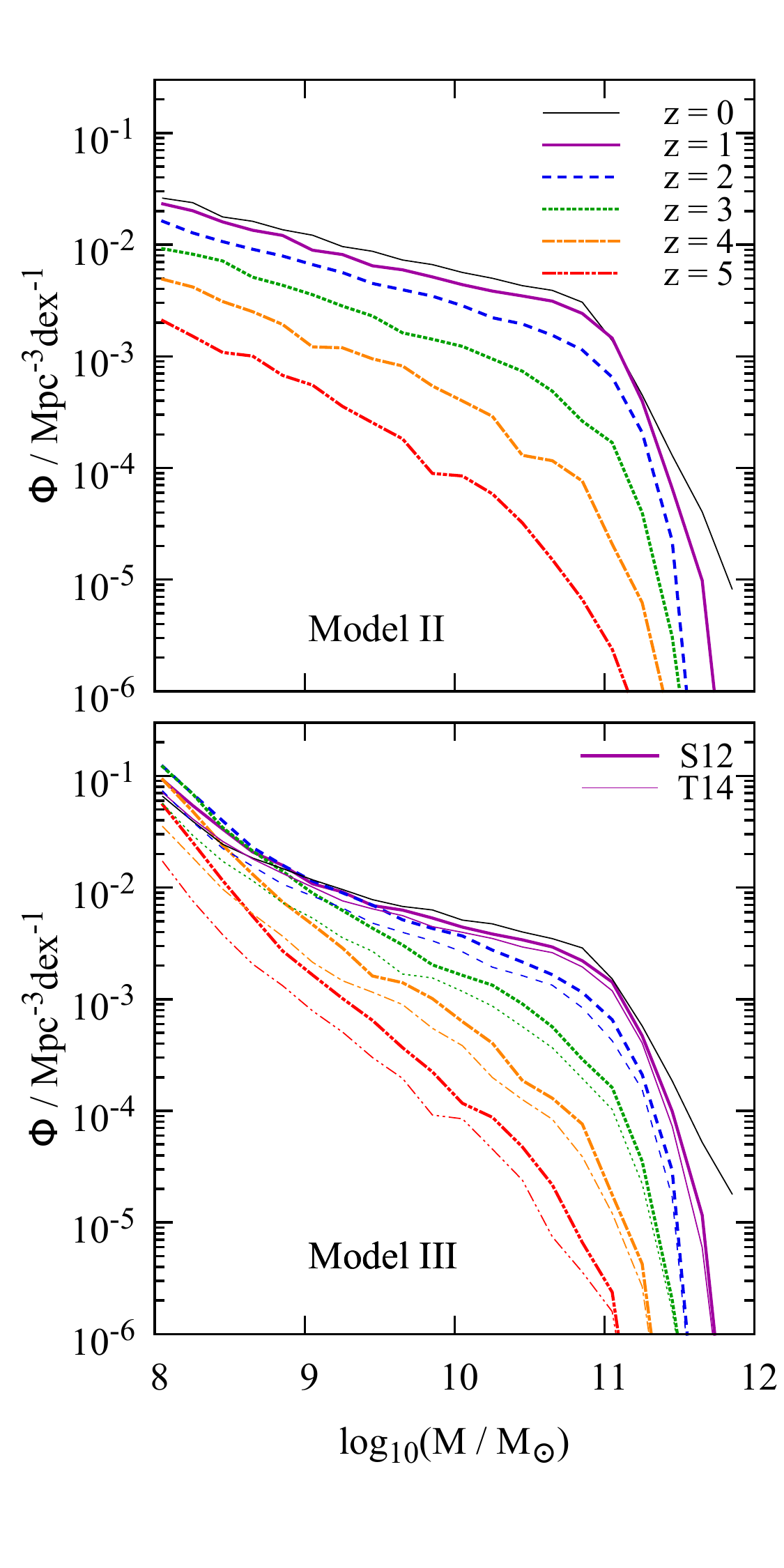}
 \end{minipage}
 \begin{minipage}{0.48\linewidth}
 \includegraphics[width=1.0\linewidth]{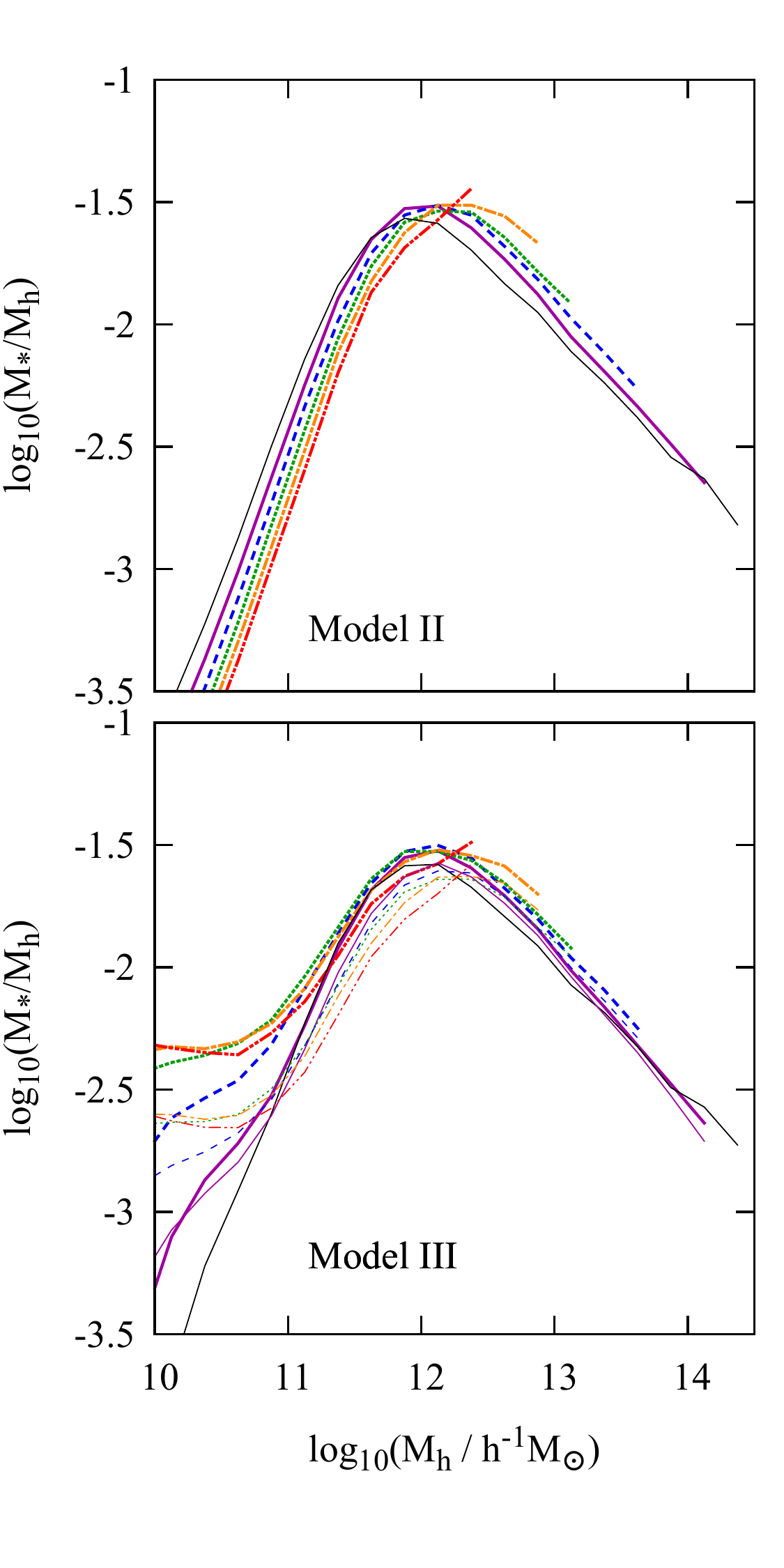}
 \end{minipage}
 \caption{Upper panels: the evolution of the SMF and 
          the $\Mstar/M_{\rm h}$ - $M_{\rm h}$ relation 
          predicted by Model II.
          Lower panels: predictions of Model III constrained
          using the SMFs from \citet{Santini12} (thick lines)
          and \citet{Tomczak14} (thin lines).}
\label{smf}
\end{figure*}

In \citet{Lu14} two sets of observational data were used to constrain
the models described above. One is the galaxy SMFs at different 
redshifts and the other is the $z$-band cluster galaxy luminosity function from \citet{Popesso06}. 
The SMFs were those from \citet{Baldry12} for $z\sim 0$, 
from \citet{PG08} for $1.0<z<1.3$, from \citet{Marchesini09} 
for $2.0<z<3.0$, and  from \citet{Stark09} for $3.19<z<4.73$. 
The results presented in \citet{Lu14} are based on 
Model II constrained by the SMFs alone and Model III 
constrained by both the SMFs and the cluster galaxy luminosity function. As mentioned above, 
Model II is unable to match the steep, faint-end upturn observed 
in the cluster galaxy luminosity function.

As shown in \citet{Lu14}, one way to distinguish Model II and 
Model III is the difference in their predicted faint-end slope 
of the SMF at high redshift. The SMF predicted by Model II 
is quite flat at the low-mass end at all redshifts, while Model III
predicts a steepening of the SMF at the low-mass end at high 
$z$. More recently, analyses based on deeper surveys provide 
some evidence for such a steepening. It is, therefore, interesting 
to use the new data to update our model parameters. 
In this paper we use two sets of SMFs recently published.
The first is that of \citet{Santini12} based on the Early 
Release Science (ERS) data of the Wide Field Camera 3 
(WFC3) in the GOODS-S Field, and the SMFs for galaxies in 
the redshift range between $0.6$ and $4.5$ are estimated 
to a stellar mass limit of a few times $10^9\Msun$. The second
set is that of \citet{Tomczak14}, who estimated the galaxy 
SMFs in the redshift range $0.2$ - $3.0$ using data from the 
FourStar Galaxy Evolution Survey (ZFOURGE) and the  Cosmic 
Assembly Near-IR Deep Extragalactic Legacy Survey (CANDELS).
Unfortunately, the two sets are not completely consistent with each other.
While at $z<2.0$ the results obtained by both are similar,  
at $z>2$ the SMFs at the low-mass ends obtained by 
\citet{Tomczak14} are lower than those obtained by 
\citet{Santini12} by a factor of $2$. In the present paper
we use the two sets of data separately to update our model 
parameters, and to check the reliability of our results against 
uncertainties in the observational data.

In practice, we replace the SMFs at $z>1$ in \citet{Lu14} with 
the new SMFs to constrain the models. As in \citet{Lu14}, we use the 
MULTINEST method developed by \citet{Feroz09}, which 
implements the nested sampling algorithm of \citet{Skilling06}, to 
explore the model parameter space and to make posterior 
predictions.
The new inferences of the model parameters are shown in 
Table~\ref{modelii} for Model II and in Table \ref{modeliii} 
for Model III. The comparison between the posterior predictions
and the constraining data is shown in Figures \ref{ppc01}
and \ref{ppc02}.
We find that without the cluster data the new data sets alone cannot 
distinguish decisively between Model II and Model III, 
because at $z>3$, the new SMFs are only complete for 
$M_{\star} > 3\times10^{9}\Msun$ where the two models 
are quite similar. The model inferences based on the \citet{Santini12}
data are similar to those obtained in \citet{Lu14}.  
The inferences from \citet{Tomczak14} data are qualitatively 
the same,  except that the inferred star formation rates
at $z\ge3$ in low-mass haloes are lower by a factor of $\sim 2$.
Note that both models over-predict slightly the SMF in the 
intermediate redshift range, and even Model III 
under-predicts the cluster galaxy luminosity function at the faint end somewhat, 
indicating that either the faint end slope in the cluster galaxy luminosity function
is overestimated, or the SMF is under-estimated in the
intermediate redshift range, or our Model III is still not 
flexible enough to match the details of the data.   

Figure~\ref{smf} demonstrates the major differences between Model II and Model III.
The upper panels show the SMFs and $\Mstar/M_{\rm h}$ 
at different redshifts predicted by 
Model II constrained by the $z\approx 0$ SMF of \citet{Baldry12} 
together with the SMFs of \citet{Santini12} at $z>1$. 
The thick lines in the lower panels show the predictions of 
Model III constrained by the $z\approx0$ SMF of \citet{Baldry12}, 
the SMFs of \citet{Santini12} at $z>1$,  and the cluster galaxy luminosity function of 
\citet{Popesso06}. As found in \citet{Lu14}, the main difference 
between Model II and Model III lies in the predicted faint end, 
especially at high redshift.  For comparison,  the thin lines 
in the lower panels show the predictions of Model III but 
constrained with the SMFs of \citet{Tomczak14}. The model constrained by the 
\citet{Santini12} data predicts systematically higher and steeper 
SMFs at $z\ge3$ than that constrained by \citet{Tomczak14}.
Clearly, accurate observations of the SMFs at $z>3$ down to 
a stellar mass limit $M_\star <10^{9}{\rm M}_{\sun}$ are crucial 
to discriminate the `Slow Evolution' Model II  and Model III advocated 
by \citet{Lu14}.  In what follows, we will use the constrained 
Model II and Model III to make model predictions for a number of 
statistical properties of the galaxy population at different
redshifts. For clarity, our presentation is based on model 
parameters constrained by the \citet{Santini12} SMFs at $z>1$, 
i.e. the values listed in the second columns of Tables 1 and 2. 
We emphasise, however, that none of our results will change 
qualitatively if the \citet{Tomczak14} data are used instead to 
constrain the models.

\section {General Trends and Characteristic Scales}
\label{sec_general}

\begin{figure*}
\centering
\includegraphics[width=0.9\linewidth]{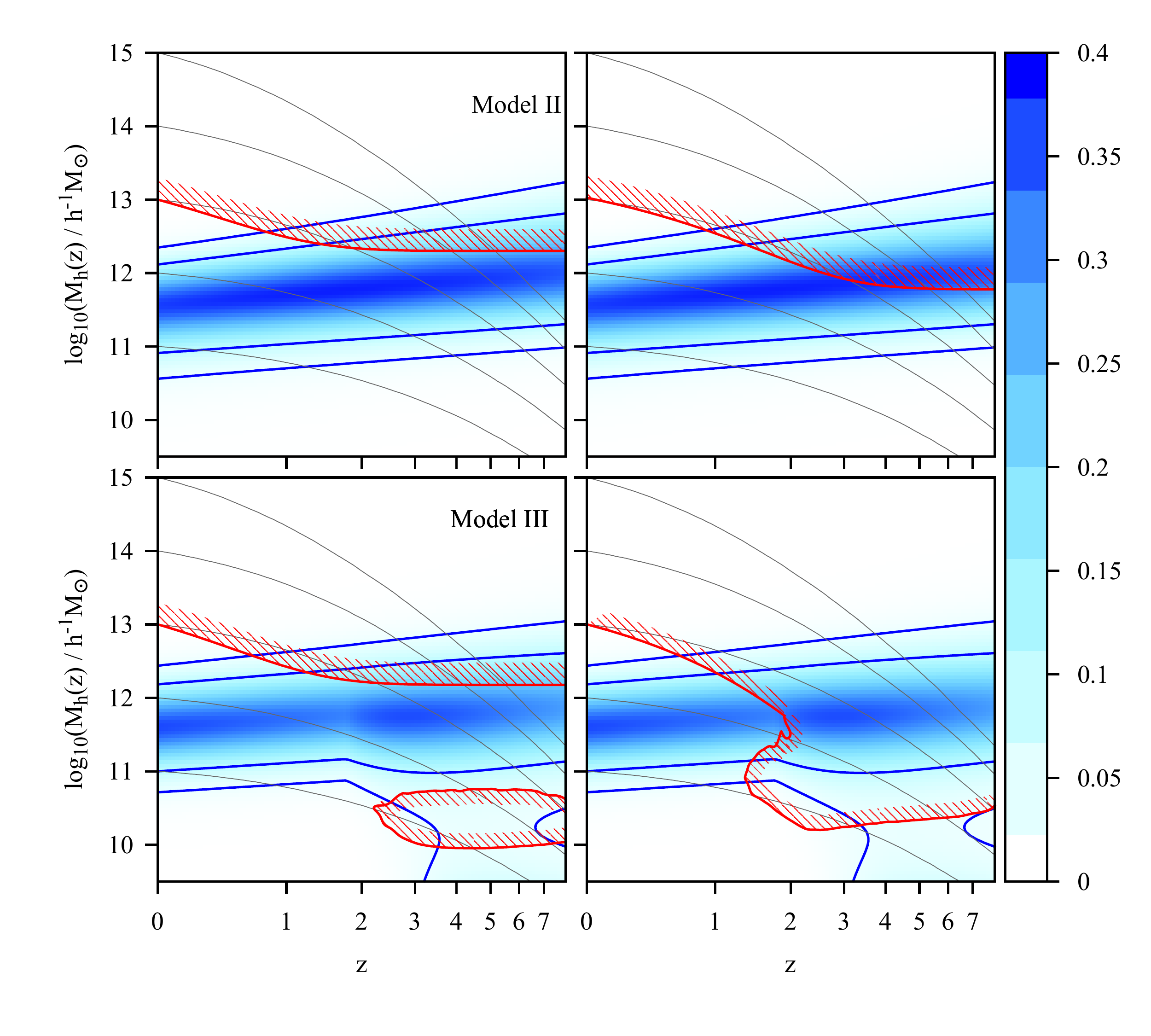}
\caption{Star formation efficiency and merger rate in the halo mass - redshift plane. The blue shading is the star formation efficiency as indicated in the
colour bar on the right.
  The blue solid lines correspond to star formation efficiency 0.1 (inner two lines) and 
  0.03 (outer lines).  The solid red lines are the loci of one merger event per Hubble time, 
  with the hedged sides corresponding to a higher merger frequency.  
  The left panels are for major mergers (with mass ratios larger than $1/3$) and the right panels are
  for minor mergers with mass ratios between $1/10$ and $1/3$.  The
  solid grey lines are the mean mass assembly histories of dark matter
  haloes with present-day masses equal to 
  $10^{11}, 10^{12}, \cdot\cdot\cdot, 10^{15}\,h^{-1}{\rm M}_\odot$, 
  which are obtained by averaging over the main branches
  of the merger trees with the same present-day halo mass}.
  The upper panels are predictions of Model II and the lower
  panels are the predictions of Model III.
\label{assembly}
\end{figure*}

Galaxies can grow their stellar mass either via 
{\it in situ} star formation or mergers.
In this section, we identify critical halo masses 
and redshifts to characterise the different stages of stellar 
mass acquisition as their host haloes grow.

A way to characterise the {\it in situ} star
formation in different haloes at different redshifts is to consider a star
formation efficiency which is defined as the ratio between 
the {\it in situ} SFR in the central galaxy and the mean halo mass 
accretion rate $\left\langle{\dot M}_{\rm h}(z)\right\rangle$ multiplied by the 
universal baryon fraction $f_B$:
\begin{equation}
\epsilon_{\rm SFR}(z)
\equiv {{\rm SFR}(z) \over f_B\left\langle{\dot M}_{\rm h}(z)\right\rangle} \,.
\end{equation}
The ${\rm SFR}(z)$, defined in Eqs\,\ref{eqn_general01} and \ref{eqn_general02},
is implicitly assumed to be an average over galaxies of similar halo mass. 
For a given halo mass $M_{\rm h}(t)$, the average mass accretion rate is 
calculated using 
\begin{equation} 
\langle {\dot M}_{\rm h}(t)\rangle 
\equiv [M_{\rm h}(t)-\left\langle M_{\rm prim}(t-{\rm d}t)\right\rangle]/{\rm d} t\,,
\end{equation} 
where $\left\langle M_{\rm prim}\right\rangle$ is the average mass 
of the primary progenitors. The primary progenitors are sampled
using the same algorithm adopted to generate the merger trees.

The plots in Figure~\ref{assembly} show $\epsilon_{\rm SFR}$ as
a function of $M_{\rm h}(z)$ and $z$. As one can see, Model II
predicts that the {\it in situ} star formation is most efficient
(with $\epsilon_{\rm SFR}>0.3$) in a narrow band between
$10^{11}\Msunh$ and $10^{12}\Msunh$, with a tilt towards lower mass
haloes at lower redshifts.  Outside the band, the star formation
efficiency drops rapidly towards both higher and lower masses without
depending strongly on redshift. This trend of star formation
efficiency with halo mass and redshift agrees with the
results obtained earlier by \citet{Bouche10}, \citet{Behroozi13b},
\citet{Bethermin13}, \citet{Tacchella13}, and \citet{Yang13}, using
different constraints and methods. Model III predicts a similar trend
except for haloes with masses $<10^{11}\Msunh$.  Instead of a strong
suppression of star formation with decreasing halo mass, the star
formation efficiency remains $\sim 1/30$ at $z>3$ 
for all $M_{\rm h}<10^{11}\Msunh$ haloes.

For central galaxies, another potentially important process that can
affect their stellar mass and perhaps the size and morphology is 
the accretion of satellites. 
To characterise this process, we calculate the mean galaxy merger rate
for haloes of a given mass at a given redshift. 
We distinguish two different types of mergers:
(i) major mergers for which the {\it stellar mass} ratio between 
the merging satellite and the central galaxy is $\ge 1/3$ and
(ii) minor mergers for which the ratio is between $1/10$ and $1/3$.  
When calculating the mass ratios, we use the original stellar mass
of the satellites before they deposit a fraction $1-f_{\rm TS}$
into the stellar halo component. 
We discuss this choice in the last paragraph of \S\ref{sec_mergers}.
We define the merger rate as the number 
of mergers per unit time multiplied by the Hubble time $t_H(z)\equiv H(z)^{-1}$.
In Figure~\ref{assembly}, the red lines are the loci of one
merger event per $t_H(z)$ in the halo mass - redshift plane; haloes on
the hedged sides of the loci on average have more than one merger per
$t_H(z)$.  We show results for both major (left panels) and minor
(right panels) mergers, and separately for Model II (upper panels) and Model III
(lower panels). 

For Model II, central galaxies at $z=0$ have
experienced at least one major merger per Hubble time if they are
hosted by haloes with masses larger than $\approx10^{13}\Msunh$. At
$z>1$, frequent major mergers only occur for centrals hosted by haloes
with masses higher than $\sim 3\times10^{12}\Msunh$. The predictions
of Model III are quite similar for massive haloes, but an additional
branch of high major merger rate is also predicted for low mass haloes
($10^{10}\Msunh < M_{\rm h} < 10^{11}\Msunh$) at $z>3$. 
Comparing the red hedged lines indicating mergers with the star
formation efficiency, one can see that most star forming galaxies 
are not associated with major mergers, 
except for central galaxies in low-mass haloes with 
$M_{\rm h}= 10^{10}$ - $10^{11}\Msunh$ at $z>3$, where galaxies 
can experience major mergers while actively forming stars. 
Minor mergers are more common. In particular, Model III predicts that
active star forming central galaxies in all haloes with
$M_{\rm h}>10^{10}\Msunh$ may have experienced at least on minor merger at $z>2$.

The following set of functions and scales
characterise the star formation efficiency and merger frequency:
\begin{itemize}
\item The ridge of the highest star formation efficiency is well
  described by
\begin{equation}
M_{\rm h} (z) \approx 3\times10^{11}\Msunh (1+z)^{0.3}\,,
\end{equation} 
with a height $\epsilon_{\rm mx}\sim 0.5$ and a FWHM $\Delta \log_{10}
(M_{\rm h})\approx 1.0$.
\item The line separating frequent from infrequent major mergers for
  massive haloes can be approximated by
\begin{equation}
M_{\rm h} (z) \approx 10^{13}\Msunh \left[0.7\exp(-z/0.6)+0.3\right]\,.
\end{equation} 
\item The line separating frequent from infrequent minor mergers in
  the entire redshift range for Model II, and at $z<2$ for Model III,
  can be approximated by
\begin{equation}
M_{\rm h} (z) \approx 10^{13}\Msunh \left[ \exp(-z/0.8)+0.06\right]\,.
\end{equation} 
\item
For Model-III, there is a characteristic redshift, $z_c\sim 2$, above which 
the star formation efficiency and the major merger frequency are boosted in low
mass haloes with $M_{\rm h} (z)\la 10^{11}\Msunh$, and the minor merger
frequency is boosted in all haloes with $M_{\rm h}(z) >10^{10}\Msunh$.
\end{itemize}

\section {Star Formation and Stellar Mass Assembly Histories}
\label{sec_histories}

\begin{figure*}
\centering
\includegraphics[width=0.95\linewidth]{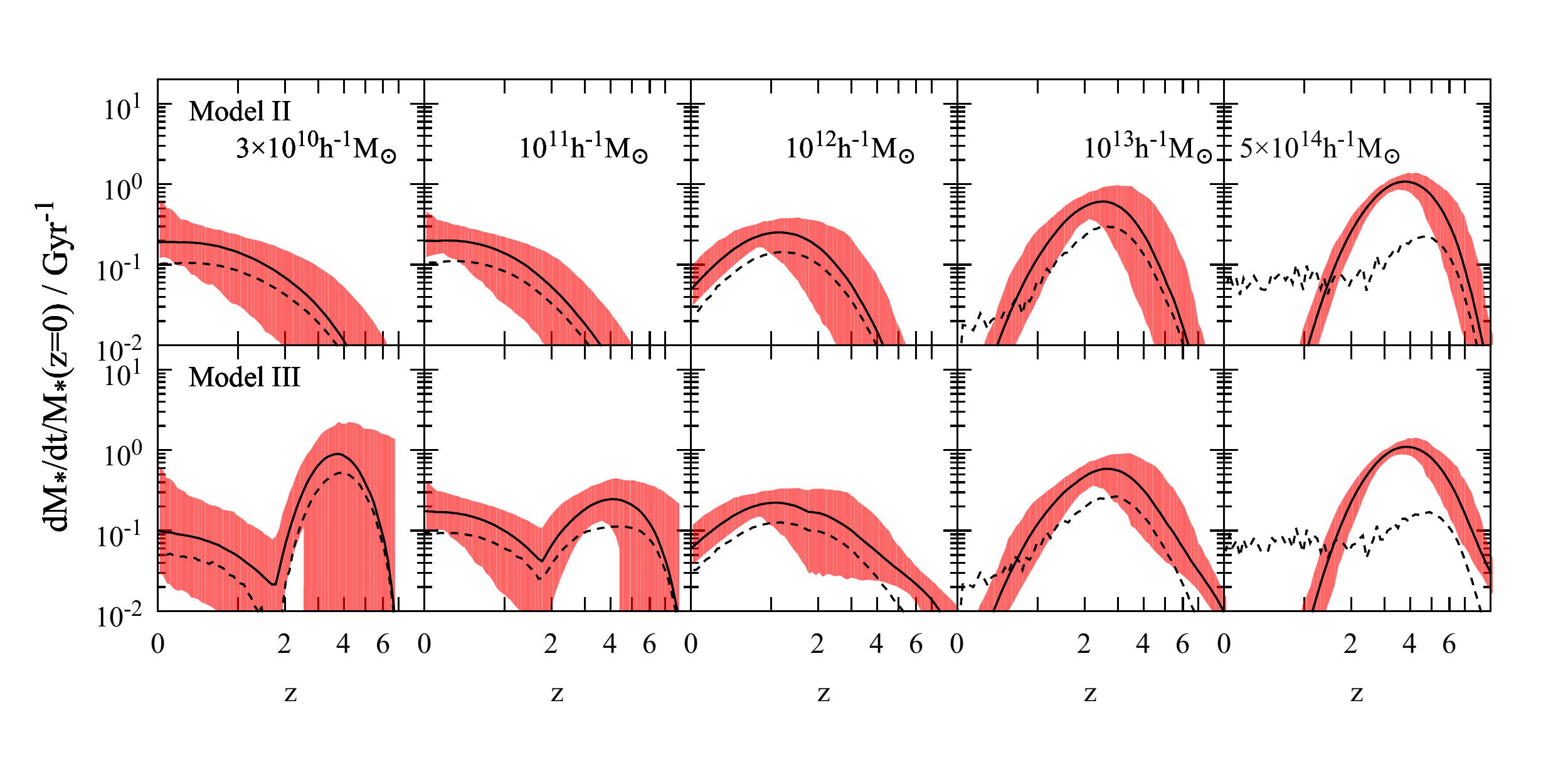}
\caption{The star formation histories in haloes with different
  present-day masses (as indicated in the upper panels)  
  predicted by the best-fit parameters  of Model II
  (upper panels) and Model III (lower panels).  The thick solid lines
  are the averages; the shaded areas are the 95\% ranges of variance
  owing to different halo merger histories.  The dashed black lines are
  the average differential assembly histories.  }
\label{fig:sfrh}
\end{figure*}

\begin{figure}
\centering
\includegraphics[width=0.9\linewidth]{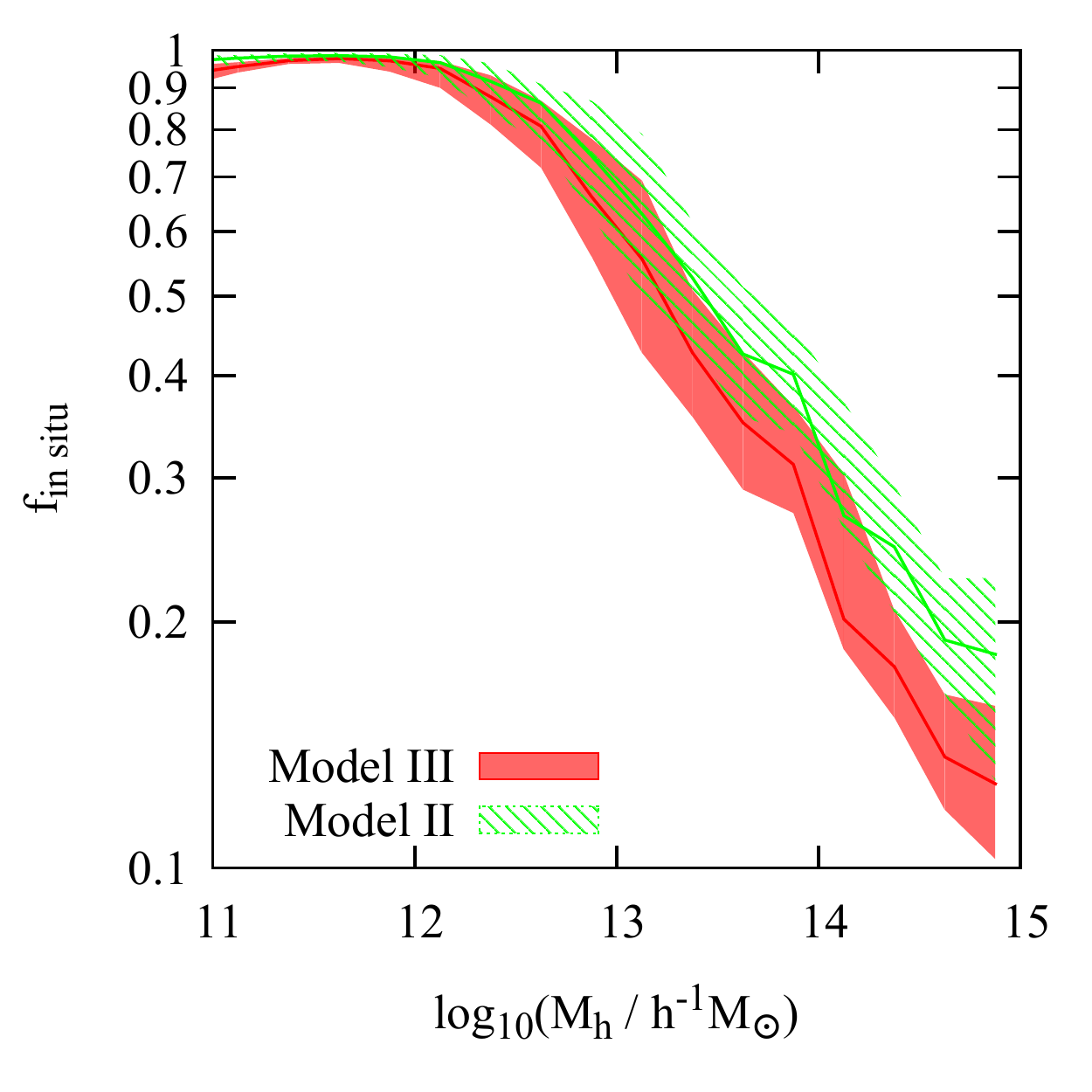}
\caption{The fraction of stars formed {\it in situ} in present day
 central galaxies of haloes with different masses. Shown are the
 predictions of Model II (green) and Model III (red). The curves are
 the averages obtained from an ensemble of halo merger trees, while
 the shaded areas represent the variances among the different
 halo merger trees.}
\label{f_z0}
\end{figure}

\begin{figure*}
\centering
\includegraphics[width=0.95\linewidth]{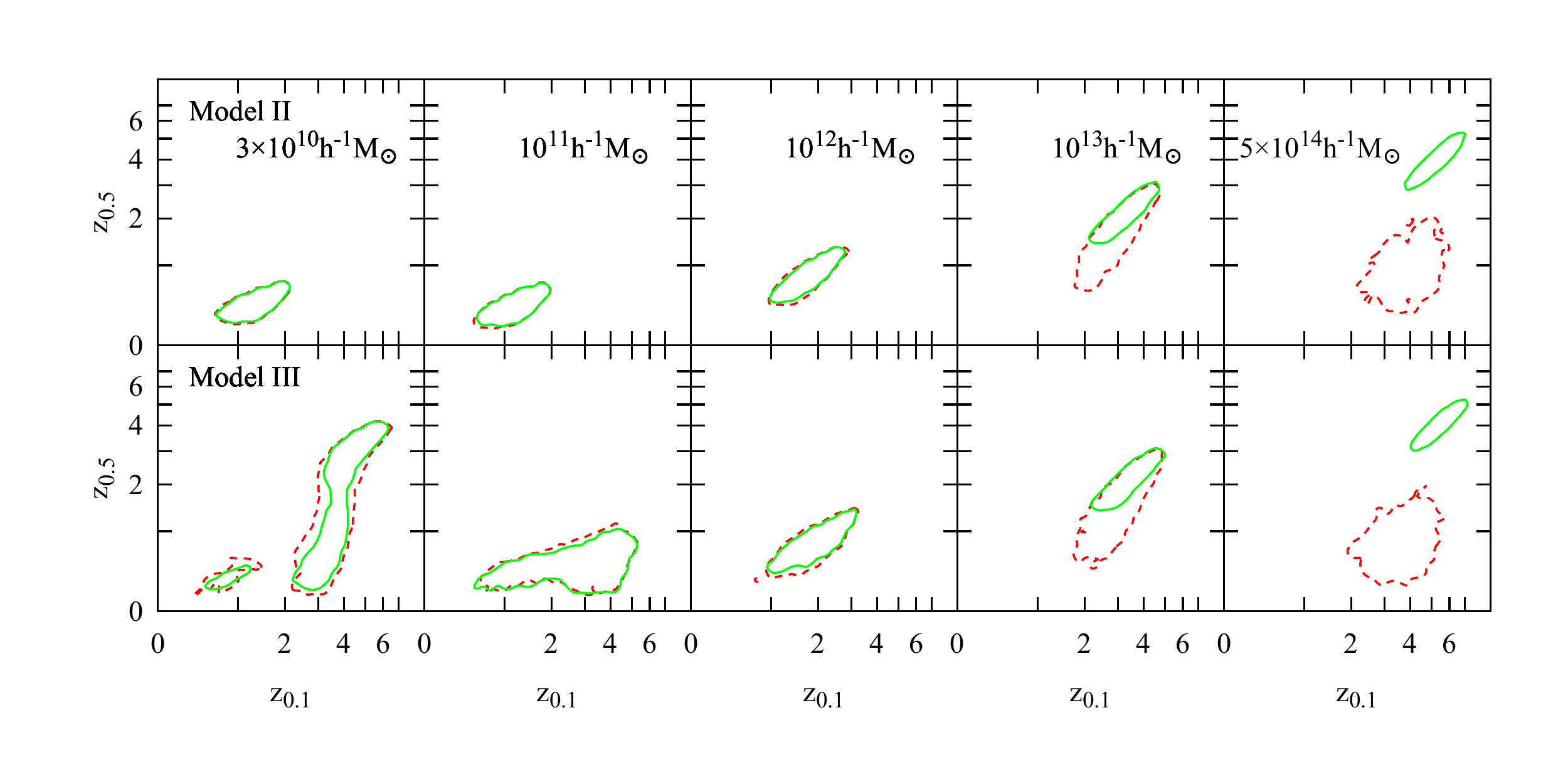}
\caption{The distribution of the redshift at which 50\% ($\zhalf$) and
10\% ($\zten$) of the stars in a central galaxy have formed
(green solid contours) and have assembled (red dashed contours).  
The contours are the isodensity lines that enclose $90\%$ of the
halo merger trees.
We show results
for haloes with five masses at $z=0$, as indicated.  
The upper panels are the predictions of Model II, 
while the lower panels are for Model III.}
\label{fig:zfdis}
\end{figure*}

\begin{figure*}
\centering
\includegraphics[width=0.8\linewidth]{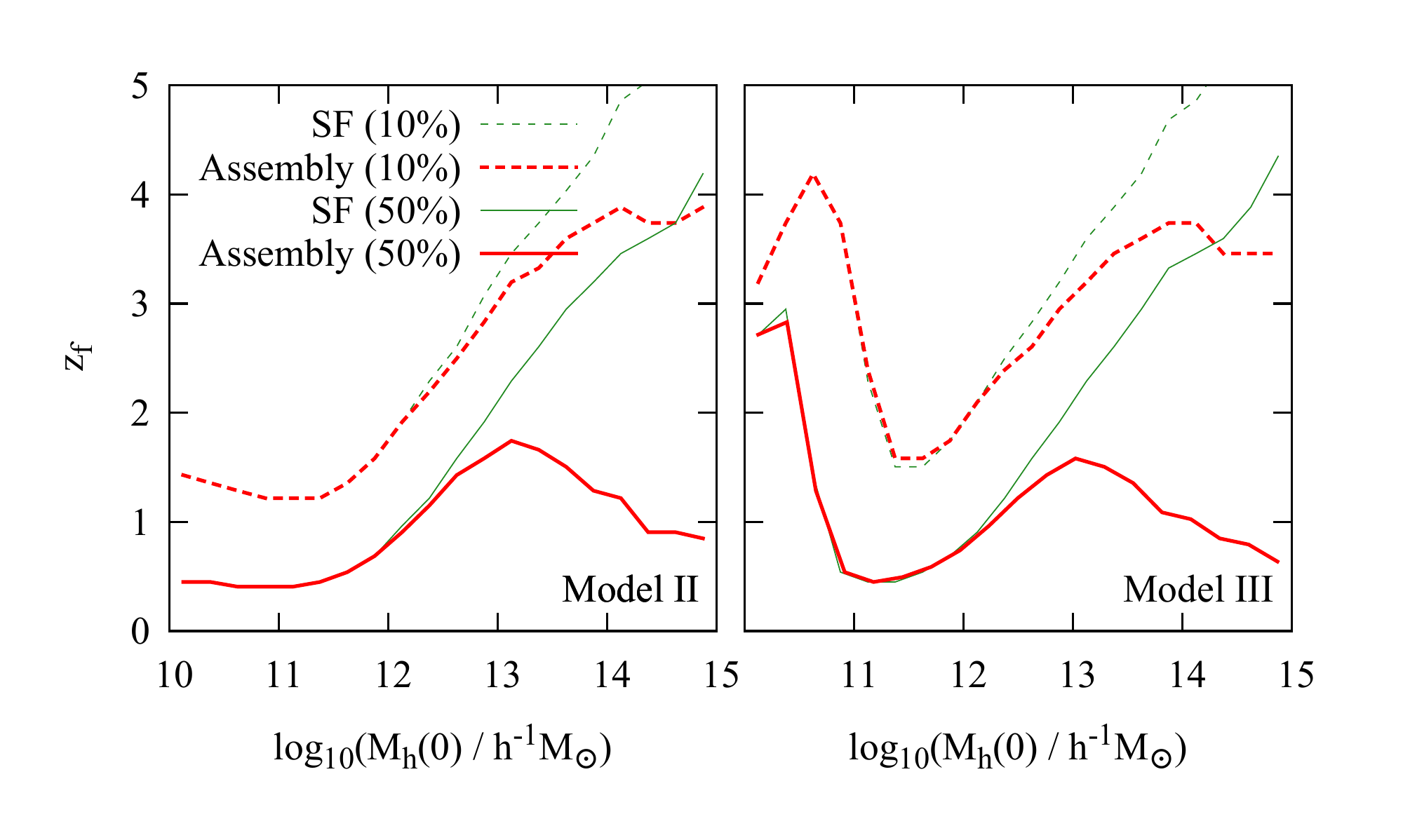}
\caption{The average characteristic redshifts,$\zhalf$ and $\zten$,
       as a function of halo mass for star formation (thin lines)
       and for stellar mass assembly (thick lines).}
\label{fig:zfdis_b}
\end{figure*}

\subsection{Star formation histories}

This total SFH (Eq\,\ref{eq:sfh}), which takes into account the 
history of the accreted stars and the stars formed {\it in situ},
is needed when modelling the stellar population of the galaxy.  
In Figure~\ref{fig:sfrh}, the black solid lines show the total SFHs 
of central galaxies in different dark halo mass ranges 
by averaging over a large number of halo merger histories
using the best fit parameters.  
For comparison, the shaded band in each panel represents the 
variance among different merger trees of the final halo mass 
in question.  We show results for both Model II 
(upper panels) and Model-III (lower panels) for five final halo 
masses, as indicated in each panel. 

It is clear that haloes of different present-day masses have different 
star formation histories.  For centrals in massive clusters
with $M_{\rm h}(z=0)\sim 5\times10^{14}\Msunh$, the SFR peaks at $z\approx 3$ and the
majority of the stars form in a narrow time range, which is about
10~Gyrs ago ($z>2$). In contrast, for Milky-Way mass haloes, the star
formation rate reaches a maximum between $z=2$ and $z=1$ and
decreases only mildly to the present day.  For dwarf galaxies in haloes
with $M_{\rm h}(z=0)<10^{11}\Msunh$, the predictions of Model II and Model III
are significantly different.  For model II, most stars in these dwarf
galaxies formed quite late (at $z\le 1$) and the SFR is roughly a constant
over this time interval. In contrast, Model III predicts star
formation histories that are bimodal, with an initial star burst at
$z>2$, followed by a constant SFR. As discussed 
in \citet{Lu14}, although the boost in the star formation rate at high $z$ in
low-mass haloes is inferred from the observed upturn in the cluster galaxy luminosity function at the 
faint end, it also has support from the observed star formation rate 
function at $z \ga 4$ \citep{Smit12}, 
and from the existence of a significant old 
stellar population in present-day dwarf galaxies \citep{Weisz11}.    

\subsection{Stellar mass assembly histories}

The average differential assembly histories are shown in
Figure\,\ref{fig:sfrh} as the dashed lines.  
The assembly history, defined in Eq.\,(\ref{eq:assembly}),
takes into account {\it in situ} star formation, accretion of 
stars already formed, and mass loss due to stellar evolution. 
The average differential evolution is obtained by first averaging 
over the assembly histories of galaxies with the same halo mass, 
and then taking the time derivative of the mean assembly histories.
For both dwarf and
Milky-Way sized galaxies, the stellar mass assembly histories are
almost parallel to the SFHs except at the beginning of star
formation. The difference in amplitude, which is about a factor of 2, 
owes to the mass loss of evolved stars.  This suggests that the assembly of such
galaxies is dominated by {\it in situ} star formation, rather than by
the accretion of stars formed in progenitors. We will have a more
detailed discussion about this in the following subsection.  For
massive cluster galaxies, the assembly histories start with a strong
episode of {\it in situ} star formation at $z>2$, which is followed by
a long period of mass accretion at roughly a constant rate.

\subsection{In-situ star formation versus accretion}
\label{ssec:insitu}

Figure\,\ref{f_z0} shows the fraction of stars formed {\it in situ} in
present day central galaxies as a function of their host halo mass.
The predictions of Model II are plotted as the green line (average) and
the green shaded area (with the variance arising from different halo merger
trees), while the predictions of Model III are plotted in red.  The
predictions of the two models are quite similar.  For central galaxies
in haloes with masses below $10^{12}\Msunh$, almost all the stars are
formed {\it in situ}.  This fraction decreases rapidly with increasing 
halo mass at $M_{\rm h}>10^{12}\Msunh$.  
About 70\% of all stars in the central galaxy of
a halo with $M_{\rm h}\sim 10^{13}\Msunh$ are formed {\it in situ}; for
cluster haloes with $M_{\rm h}\sim 10^{15}\Msunh$ this fraction is about 15\%,
so about $85\%$ of the stellar mass is acquired through accretion. 

\subsection{Downsizing versus upsizing}

Galaxies of different masses have different star formation
(assembly) histories. To characterise these histories in a
more quantitative way, we examine the characteristic redshift, $z_f$, 
by which a fraction $f$ of the final stellar mass in a galaxy has
formed (or assembled).  Figure~\ref{fig:zfdis} shows 
the distribution of galaxies in the $\zhalf$ - $\zten$ plane for
star formation (green contours) and stellar mass assembly (red
contours), with the contours delineating the isodensity lines that
contain $90\%$ of all the galaxies.  The results are again shown for haloes
with five different present-day masses, as indicated in the panels, predicted by
Model II (the upper 5 panels) and Model III
(the lower 5 panels), respectively. 
For the most massive haloes [$M_{\rm h}(z=0)>10^{14}\Msunh$],
the star formation time and
the assembly time differ considerably, especially in $\zhalf$. On average
about 50\% of the stellar mass in the central galaxies of such massive
haloes form before $z=4$, but assemble much later at $z\approx 1$.
For haloes with masses lower than $10^{13}\Msunh$, the star formation
time and assembly time are almost identical, indicating that central 
galaxies in such haloes acquire their stars mostly through {\it in situ} star
formation, as we have already seen in the last subsection. 
A Milky Way mass galaxy [$M_{\rm h}(z=0)\sim 10^{12}\Msunh$] 
on average formed about 10\% of its
stars by $z\approx2$ and about 50\% after $z=1$. For dwarf galaxies
residing in haloes with $M_{\rm h}(z=0)< 10^{11}\Msunh$, Model III predicts
diverse formation redshifts. For example, the majority of galaxies
residing in $10^{11}\Msunh$ haloes are predicted to form $10\%$ of their
stars by about $z=4$, while a fraction is predicted to form their first $10\%$
at much later times: $z\approx1$. The diversity becomes larger for
smaller galaxies: for galaxies in $3\times10^{10}\Msunh$ haloes, some show
star formation as early as that in the centrals of galaxy clusters, 
while others formed most of their stars after $z\approx1$.
This diversity owes to the transition in star formation 
efficiency at $z\approx2$ and the variance in the halo accretion histories. 
Haloes that formed early generally have experienced an early burst
phase of star formation, while younger haloes that
assembled most of their mass later than the transition redshift
did not have such an early star burst.

In Figure\,\ref{fig:zfdis_b} we show how the
averages of $\zten$ and $\zhalf$ change with halo mass for both star
formation (thin lines) and stellar mass assembly (thick lines).  For
haloes with $M_{\rm h}(z=0)>10^{11}\Msunh$, both Model II and Model III predict
that the centrals of more massive haloes on average form a fixed
fraction of their stars earlier, a trend usually referred to as
``downsizing'' \citep{Fontanot09, Weinmann12}. 
A similar downsizing trend is also seen in the
stellar mass assembly for haloes with $10^{11}\Msunh<M_{\rm h}(z=0)<10^{13}\Msunh$.  
However, this trend breaks down in two ways.  
First, for the most massive haloes [$M_{\rm h}(z=0)>10^{13}\Msunh$], more
massive centrals actually assemble their stars later. 
At late times, these galaxies tend to build up their mass
hierarchically by accreting satellite galaxies, resembling the
mass assembly history of the dark matter haloes themselves.  
Second, although the
downsizing trend holds for dwarf galaxies in Model II, the trend
predicted by Model III is completely the opposite for these galaxies:
on average smaller galaxies tend to be older. This again owes to 
the boost of star formation in low-mass haloes at high $z$ in 
Model III.
 
\subsection{The global star formation history}
\begin{figure}
 \centering
 \includegraphics[width=0.9\linewidth]{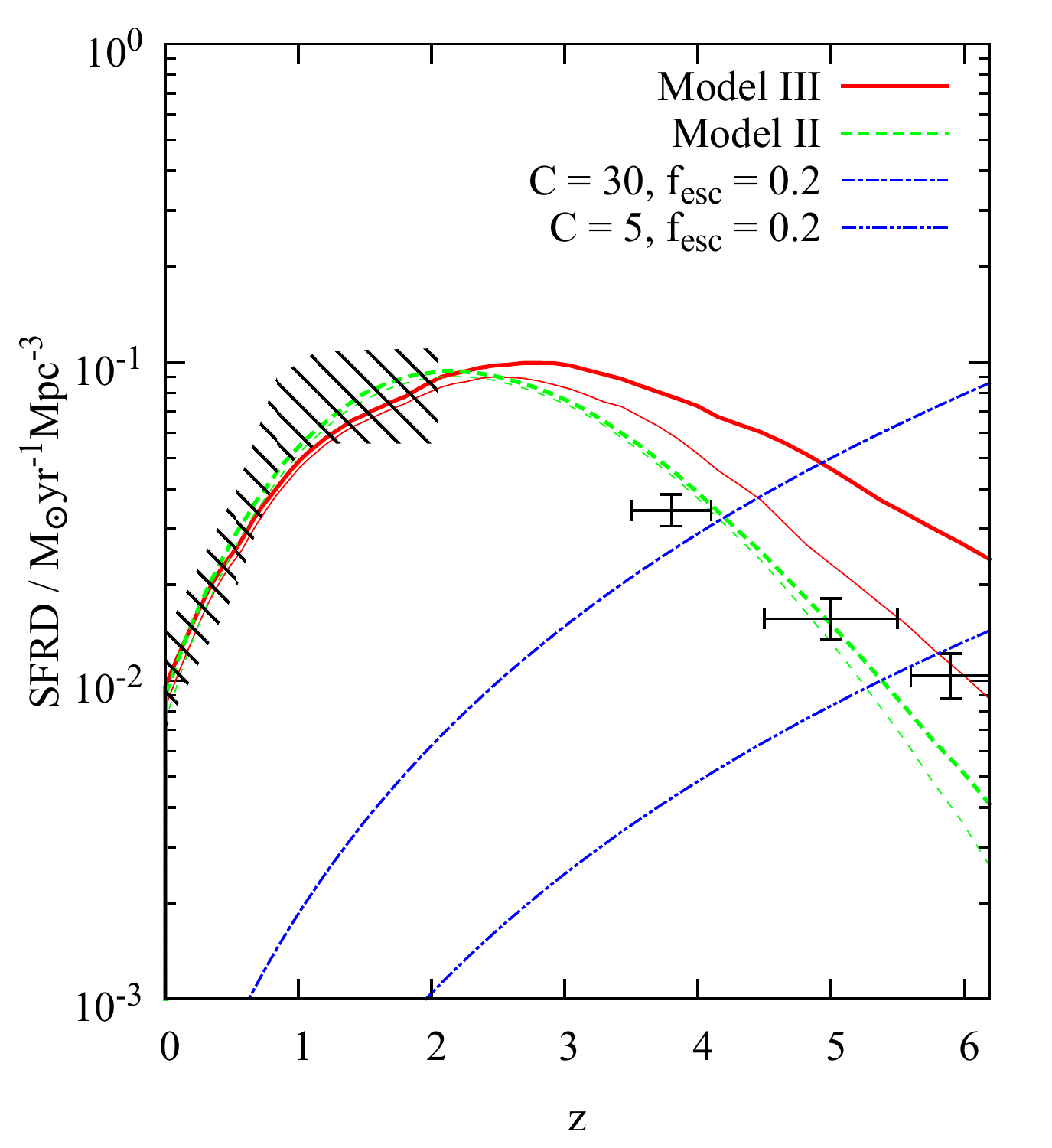}
 \caption{
   The predicted star formation rate density (solid and dashed lines) 
   and the minimal star formation rate density 
   required to keep the Universe ionised (dot-dashed lines).
   The hedged band and the data points are from \citet{Bouwens12}.
   For the model predictions, the solid lines are obtained by integrating
   down to the merger tree resolution, which is $2\times 10^{9}\Msunh$,
   and the dashed lines are obtained by integrating down to the detection
   limit in \citet{Bouwens12}, which is $M_{\rm UV, AB} = -17.7$ or
   $0.36M_{\odot}{\rm yr}^{-1}$.}
 \label{reionize}
\end{figure}

The solid green and red lines in Figure\,\ref{reionize} show the
star formation rate density (SFRD) predicted by Model II and Model III,
respectively. These results take into account star formation 
in haloes down to a mass of $2\times 10^{9}\Msunh$. 
As shown in \citet{Lu14}, the boost in the SFRD at $z>3$ of
Model III relative to Model II is dominated by star formation in
low-mass galaxies hosted by haloes with masses $M_{\rm h}<
10^{10.5}\Msunh$. These galaxies are missed in the current
observational data used to derive the SFRD, and so the discrepancy
between the prediction of Model III and the observational results 
at $z>3$ (shown as error bars in the figure) 
probably owes to incompleteness in the data. Indeed, if we use the 
same lower limit of UV magnitude adopted in \citet{Bouwens12},
which is $M_{\rm UV, AB} = -17.7$, to predict the SFRD, 
we get the results shown by the thin lines, 
which brings the prediction of Model III into much better agreement 
with the data. The change 
in the prediction of Model II is small, because in this model 
galaxies below the limit do not make a significant contribution.
This demonstrates clearly the importance in observing and modelling 
very faint galaxies to understand the SFRD at high $z$.

If one assumes that the Universe is kept ionised by the UV photons
from young stars, a minimal SFRD required can be estimated using 
the equation given by \citet{Madau99}:
\begin{equation}
{ {\rm SFRD}_{\rm min}(z)\over
{\rm \Msun yr^{-1} Mpc^{-3}}}
 \approx 3\times10^{-4}
 \left(\frac{C}{f_{\rm esc}}\right) 
\left(\frac{1+z}{6}\right)^3 \,, 
\end{equation}
where $C$ is the clumpiness factor of the IGM and $f_{\rm esc}$ is the
average escape fraction of UV photons from star forming galaxies. The
normalisation factor is consistent with the Chabrier IMF \citep{Chabrier03}. 
The clumpiness factor is expected to evolve with redshift and its value 
is quite uncertain. \citet{Madau99} adopted $C=30$ at $z=5$ based on 
the cosmological simulation of \citet{Gnedin97}. 
More recently, \citet{Bolton07} claimed that $C$ at
$z=7$ may be as small as 5.  For comparison we show in
Figure\,\ref{reionize} the minimum SFRD as a function of $z$ assuming
$C=30$ and $C=5$, with $f_{\rm esc}=0.2$ in both cases. 
If the clumpiness factor is as high as 30,
Model III is able to keep the Universe ionised at $z\approx5$, while
SFRD of Model II is too low. However, without better constraints on 
$C$ and $f_{\rm esc}$,  no stronger statement can be made about 
the two models.

\section{Merger history and the transformation of galaxies }
\label{sec_mergers}

\begin{figure}
\centering
\includegraphics[width=\linewidth]{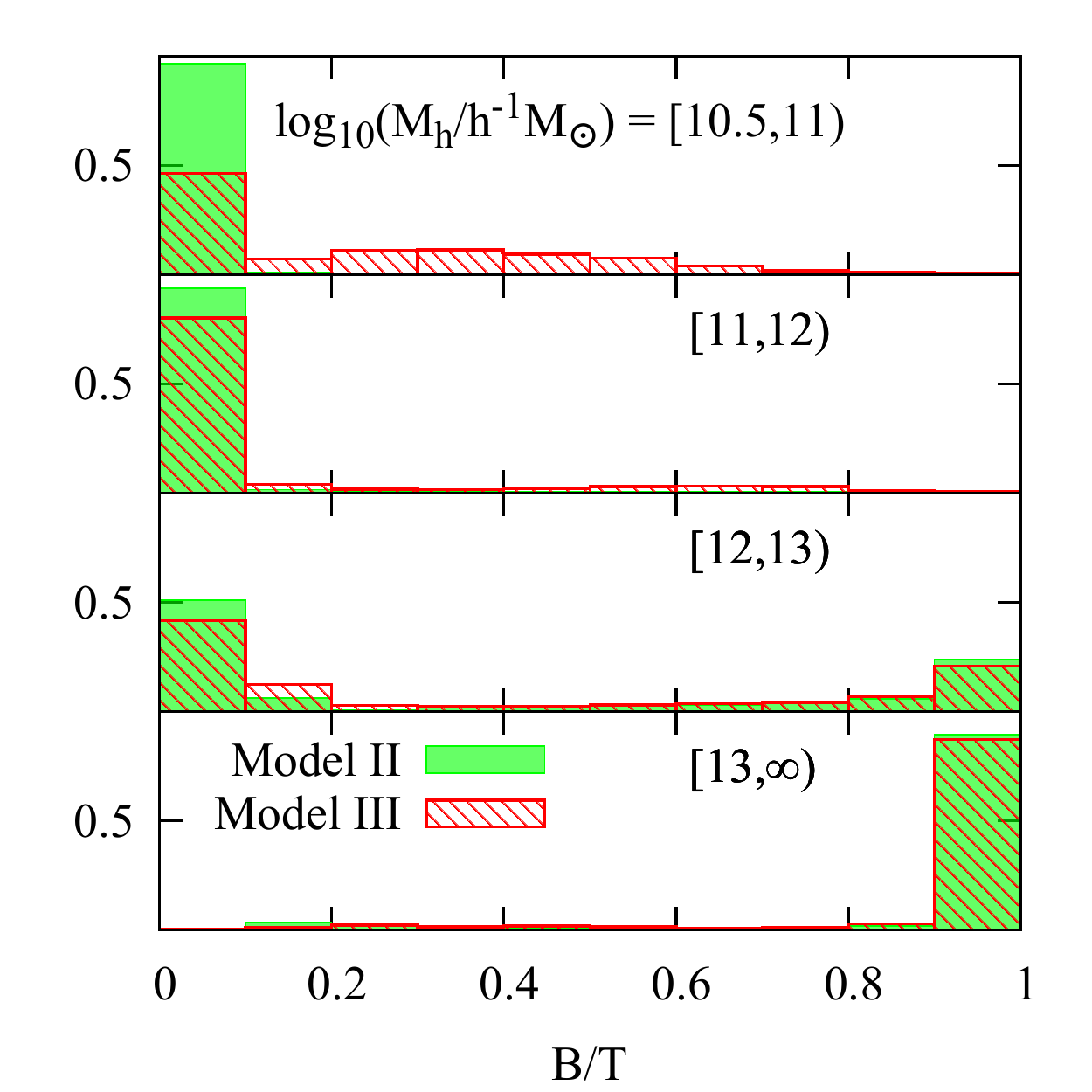}
\caption{The bulge-to-total mass ratio of local central galaxies as a
  function of host halo mass. The green histograms are the predictions of
  Model II while the red hedged histograms are those of Model III.}
\label{bulge}
\end{figure}

\begin{figure}
\centering
\includegraphics[width=\linewidth]{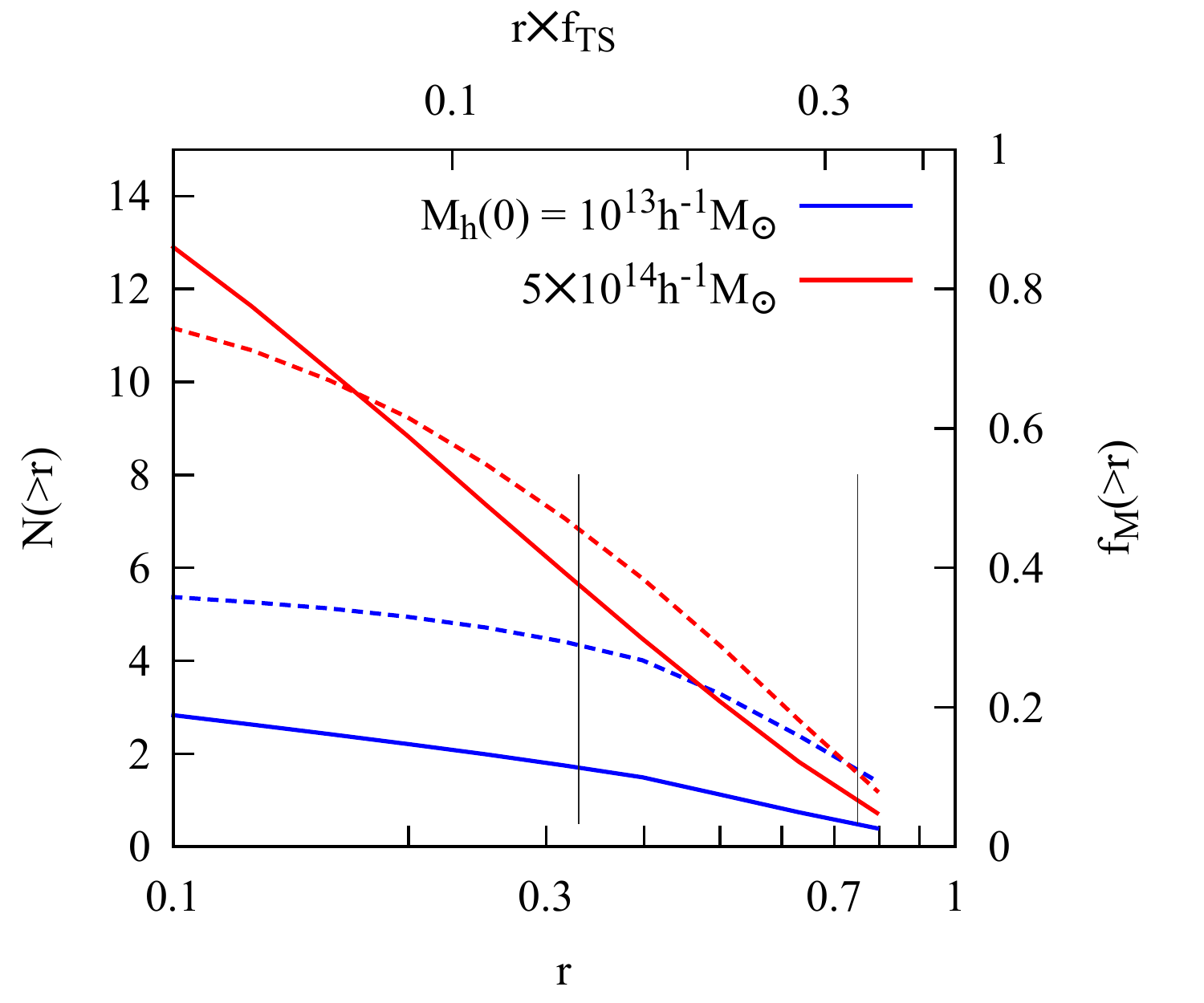}
\caption{Solid curves: the number of merger events since the {\it in situ}
         star formation rate declines to $1/3$ of the peak value
         with a mass ratio $>r$, where $r$ is the stellar mass 
         ratio between the secondary and primary merging progenitors.
         Dashed curves: the fraction of the total stellar mass acquired
         by mergers with a mass ratio $>r$.
         The left vertical line marks $r=1/3$, which separates 
         major from minor mergers. The results are shown for Model III
         (the results for Model II are similar for the massive 
         haloes in question). For comparison we also label 
         $r\times f_{\rm TS}$ on the top of the figure, and mark 
         $r\times f_{\rm TS}=1/3$ by the vertical line on the right.
         Since $f_{\rm TS}$ is the fraction of the satellite mass
         that ends up in the central, $r\times f_{\rm TS}=1/3$ 
         represents that ratio between the satellite mass that 
         actually ends up in the central and the central mass. 
}
\label{mergercnt}
\end{figure}

\begin{figure}
\centering
\includegraphics[width=\linewidth]{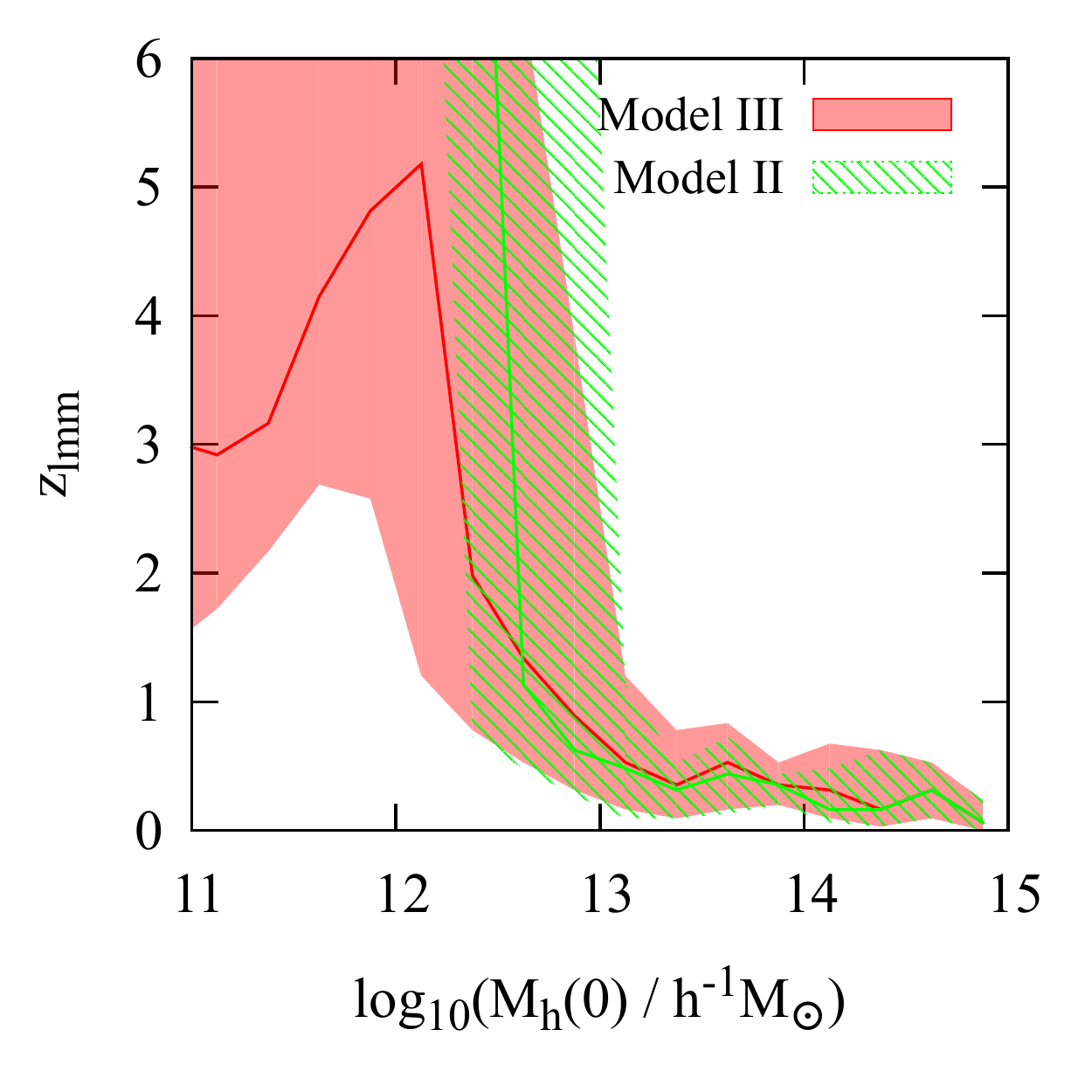}
\caption{The redshift of the last major merger of present-day central
 galaxies as a function of host halo mass. The predictions of Model II
and Model III are shown in green and red, respectively. Lines are the 
averages, while the shaded areas enclose the 95\% range of variance owing 
to different merger histories.}
\label{zlmm}
\end{figure}

In this section we characterise the galaxy merger histories in more details
and use a simple model to study their implications for the morphological 
transformation of galaxies.  

We make the simple assumption that stellar disks form only through 
{\it in situ} star formation, and that a major merger can transform 
a stellar disk into a spheroid.
Here, major mergers are defined in the same way as in \S\ref{sec_general}.
The mass of the stellar disk is then 
simply the total mass of  stars formed {\it in situ} after the last major merger.
It should be cautioned, however, that a galactic
bulge can be formed in other ways, such as the secular evolution of the disk.
The bulge mass defined here, therefore, can only be
taken as a lower limit and serves as a simple indicator as 
to how a central galaxy may be disturbed by infalling satellites.
Figure~\ref{bulge} shows the distribution of `bulge-to-total' ratio, $B/T$, 
for present-day central galaxies in four halo mass bins. 

In cluster haloes, the central galaxies are identified as BCGs in
observations. They have been bombarded quite frequently 
by satellites of different masses. As shown in  Figure~\ref{mergercnt}, 
such galaxies on average have experienced about $5$ major mergers 
and an even larger number of minor mergers during the last period of low {\it in situ} 
star formation (defined, quite arbitrarily, as the period after 
the {\it in-situ} star formation rate declines to be $1/3$ of the peak value).
Major mergers contribute about $50\%$ of the total stellar mass.
This casts doubt on the scenario in which the late growth and structure 
of BCGs are assumed to be determined by minor mergers 
\citep[e.g.][]{Bezanson09,vanDokkum10,Hilz13}.
The last major mergers happened quite recently, at $z\sim 0.5$,
as shown in Figure\,\ref{zlmm}. By inspecting images of BCGs at 
low $z$, \citet{McIntosh08} and \citet{Tran13} 
did find that a significant fraction of them indeed show signatures of recent 
major mergers, consistent with our predictions.

For haloes in the mass range $10^{12}\Msunh$ to $10^{13}\Msunh$
there is strong bimodality in their $B/T$ distribution.
Significant mergers are sparse in their entire histories:
the `ellipticals' in such low mass groups on average only experienced 
$\approx1.5$ major mergers and no more than $3$ mergers with a 
mass ratio $>0.1$ (see Figure~\ref{mergercnt}).
The central galaxy is either dominated by bulge,
if it has experienced a recent major merger, or remains disk
dominated, if such a merger did not occur. 

For a Milky Way sized galaxy with
$M_{\rm h}(0)= 10^{11.5}$-$10^{12.5}\Msunh$, {\it in situ} star formation
has dominated, while major mergers have been rare since $z\sim
2$. These galaxies, therefore, remain disk dominated, free of any 
significant major merger - driven bulge components.  

\begin{figure}
\includegraphics[width=.98\linewidth]{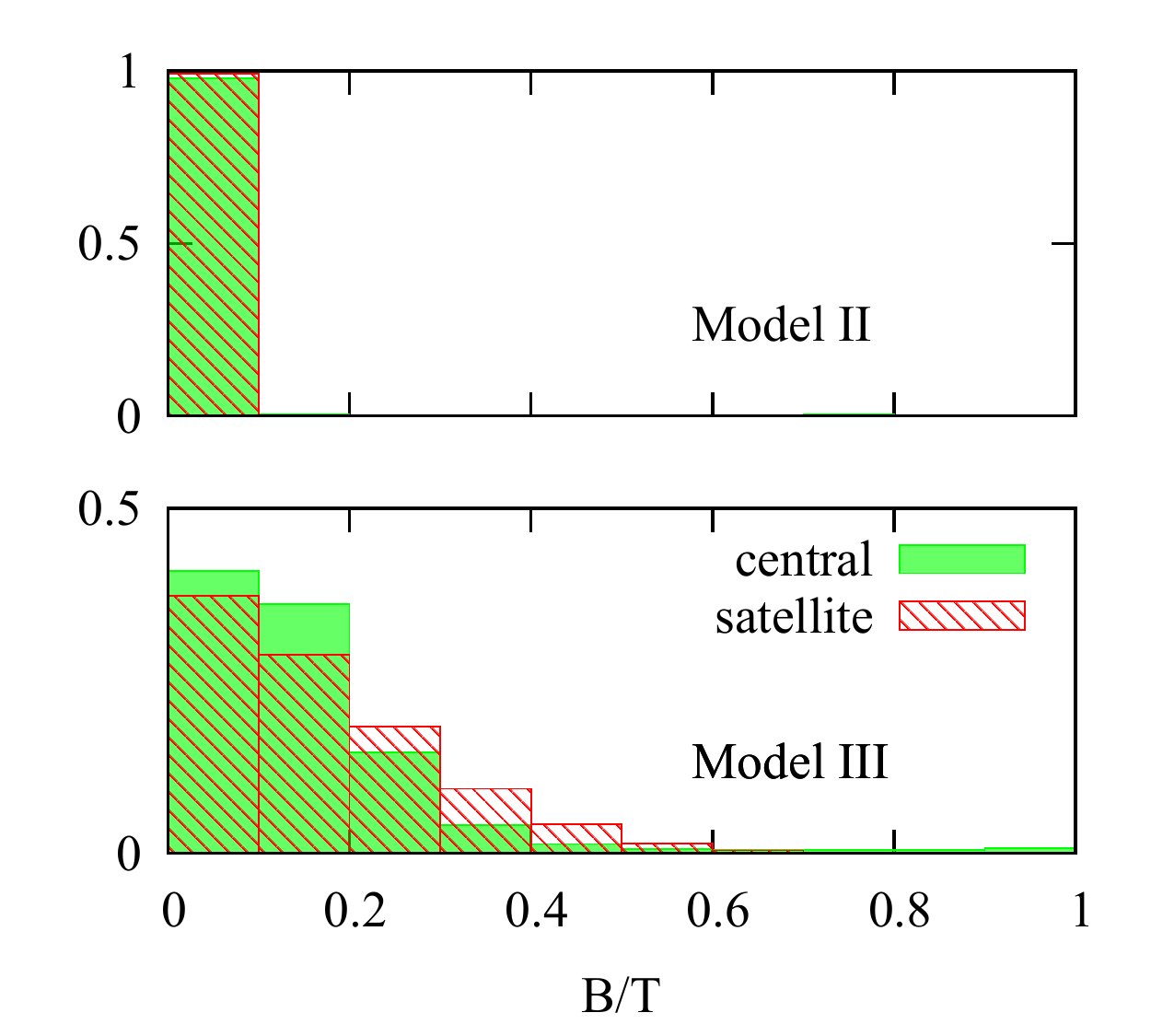}
\caption{The ``bulge'' mass fraction of dwarf galaxies with total
  stellar masses in the range $10^{8}\Msunhh$ to $10^{9}\Msunhh$,
  predicted by Model II (upper panel) and Model III (lower panel)
  for centrals (green) and for satellites (red). }
\label{bulge_dw}
\end{figure}

The predictions of Model II and Model III differ when it comes to dwarf galaxies. 
Model II, consistent with many other similar models in the literature, indicates 
major mergers between dwarf galaxies are extremely rare. Therefore,
all the stars are expected to remain in a disk (see the upper panel
of Figure\,\ref{bulge}). In contrast, Model III predicts that
most of the galaxies experienced some major mergers
during their initial star burst phases ($z>2$). The exact time
when a major merger occurs has large variations for such galaxies, 
as shown  in Figure~\ref{zlmm}.
Major mergers are very rare at $z<2$ while {\it in situ} star formation
continues, allowing the growth of new disks. The fraction of
stars contained in the spheroid depends both on when the last major
merger occurs and on the {\it in situ} star formation that follows.  
This complexity in star formation history results in diverse
morphologies of present day dwarfs, as shown in Figure~\ref{bulge}.

Major mergers at early stages of the evolution may shed light
on the origin of dwarf ellipticals (dE) and dwarf spheroidal (dSph) galaxies.
One popular scenario is galaxy harassment \citep{Moore96}, 
in which high speed encounters of a dwarf disk with other galaxies 
in a dense environment heats up the disk and
transform it into a dE or a dSph. However, the predicted kinematics,
which shows significant rotation, is at odds with the
the observational results \citep[e.g.][]{Geha03, Toloba13}. 
Our result here suggests that some of the slow rotators could be the 
remnants of early major mergers between dwarf galaxies. 
The lower panel of Figure~\ref{bulge_dw} shows $B/T$ of both dwarf centrals 
and satellites predicted by Model III, with the latter skewed towards higher $B/T$. 
In contrast, Model II indicates that all dwarf galaxies are strongly disk dominated. 
However, the predicted fraction of bulge-dominated dwarfs by Model III may be too 
low to account for the total population of dE's and dSph's.  
It is likely that later evolution, such as harassment, also plays a 
role in transforming the disk component, making the bulge components 
more dominant. 

It should be pointed out that the above discussion is based 
on the value of $r$, namely the ratio between the original 
secondary mass and the primary mass. As shown above, 
our model predicts that only a fraction, $f_{\rm TS}\sim 0.4$, 
of the original mass of the secondary is added to the 
central after a merger while the rest is stripped and 
deposited as halo stars. If the stripping occurred at large 
distances from the central, the secondary mass at the time of
merger would be $r\times f_{\rm TS}$. For comparison, 
the values of this quantity are labelled on the top of  
Figure\,\ref{mergercnt}. As is clear, in this case major mergers,
still defined by a mass ratio $\ge 1/3$,  would be rare 
for all galaxies. However, this scenario may not be realistic. 
Observations show that halo stars are mostly identified 
around central galaxies and, as we will see below, the 
amount of halo stars predicted by our models is 
consistent with observations \citep{Bernardi13}. This suggests 
that stripping of stars most likely occurs when the secondary 
is close to the primary, and so the mass ratio, $r$, defined 
above is more relevant when considering the mutual 
gravitational interaction between the merging galaxies. 

\section{Stellar populations}
\label{sec_starpopulation}

\subsection{Stellar ages}

\begin{figure}
 \centering 
 \includegraphics[width=\linewidth]{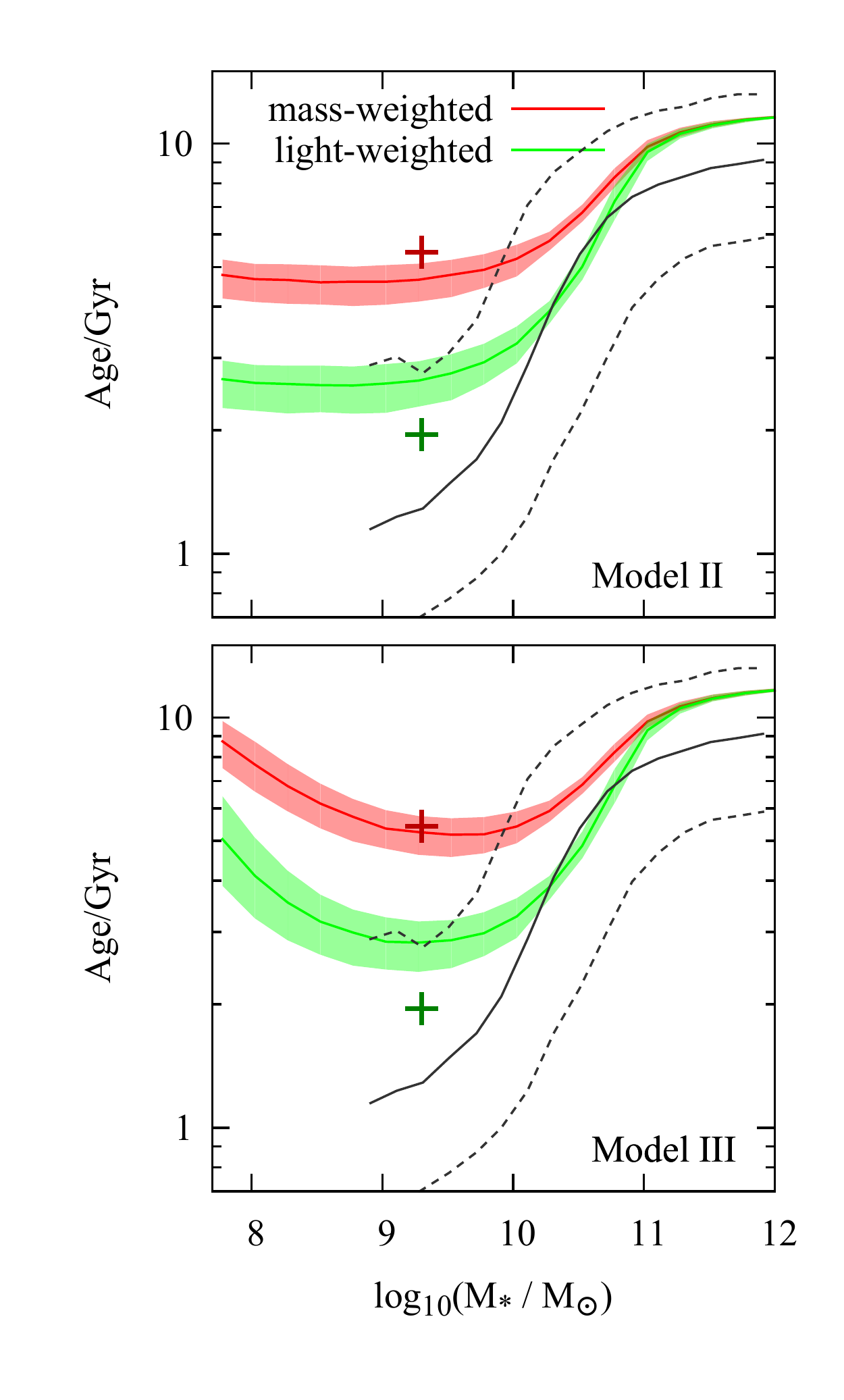}
 \caption{The stellar age as a function of stellar mass predicted by
Model II (upper panel) and Model III (lower panel). Both mass-weighted 
and light (luminosity) -weighted averages are shown, with the bands 
representing the 95\% percentile of halo merger histories. The lines 
show the light-weighted averages of stellar ages obtained by 
\citet{Gallazzi05} from the SDSS, with the solid line being the 
median and the two dashed lines representing the $16\%$ and $84\%$ percentile.    
The crosses are the mass-weighted (dark red) and
light-weighted (green) ages of the LMC obtained from the 
star formation history given by \citet{Weisz13}.
}
\label{age}
\end{figure}

Observationally, the star formation histories (SFHs) of individual
galaxies can be estimated from their stellar populations.  One popular
way to do this is to use the spectra of galaxies, which contain
information about the SFH, initial mass function (IMF), chemical
enrichment history and dust content of galaxies. In Figure~\ref{age}, 
we compare the average stellar ages predicted by Model II and Model III
with the observational results of SDSS galaxies obtained by
\citet{Gallazzi05}. The observational stellar ages are determined
by simultaneously fitting five spectral absorption features that 
break some degeneracies in spectral synthesis parameters.  
The weighted age of a galaxy is defined as 
\begin{equation}
 {\rm Age} = \frac{\int_{0}^{\infty}{\rm SFR}(t)f(t)t\,dt}{\int_{0}^{\infty}{\rm SFR}(t)f(t)\,dt}\,, 
\end{equation}
where $f(t)$ is either the fraction of the remaining stellar mass 
or the luminosity of a simple stellar population with age $t$.  In
Figure~\ref{age} we plot the average ages weighted either by 
stellar mass (red bands) or by the $r$-band luminosity (green bands). 
The observational ages, with the median shown by the solid line
and the $16$ and $84$ percentile by the dashed lines, are also weighted 
by the $r$-band luminosity, so a direct comparison between the model 
and the data can be made. The SDSS galaxies cover a stellar mass range 
between $10^{9}$ and $10^{12}\Msun$, where the observed age 
increases with stellar mass generally, and increases sharply around 
$10^{10}\Msun$. The model predictions agree with the observational
data only qualitatively: the predicted ages are systematically older
than that observed between $10^9\Msun$ and $10^{10}\Msun$. 
Similar discrepancies have also been found by \citet{LY13} using
semi-analytic models.

This discrepancy may indicate an intrinsic deficiency 
in the approach adopted here or in similar approaches in the
literature. By matching only SMFs of galaxies, our model may 
be insensitive to the star formation history in the recent past. 
For example, an enhancement in recent star formation may 
contribute little to the total stellar mass, and so is not well
captured in our model, but can significantly increase the optical
luminosity, thereby decreasing the light-weighted age. As a
demonstration, the upper cross in Figure\,\ref{age} shows the 
stellar mass weighted age of stars in the LMC based on the star 
formation history derived from the colour magnitude diagram of resolved stars
\citep{Weisz13}. Our model predictions match the observation 
well.  For comparison, the lower cross shows the $r$-band 
luminosity-weighted age obtained from the same star formation 
history with our adopted spectral synthesis model. The age so
obtained lies below the green band, because our average model 
underpredicts the current star formation rate of the LMC.        

There are uncertainties in the observational data too. 
It is in general difficult to distinguish stars that formed about 
8-10 Gyrs ago from those that formed 4-5 Gyrs ago from 
an analysis of the optical-NIR  spectra \citep[e.g.][]{Bruzual03,Pacifici12}. 
Basically the UV light provides information about recent star formation, while 
strong Balmer absorption lines are sensitive only to 
intermediate-age (~1-3 Gyrs old) stars, but one cannot distinguish 
stars older than 4 Gyrs.
Thus, it is possible that the average stellar ages derived from the 
observational data have missed the contribution of such an old 
population.
  
\subsection{Halo Stars}
\label{sec:halostars}

\begin{figure}
\centering
\includegraphics[width=\linewidth]{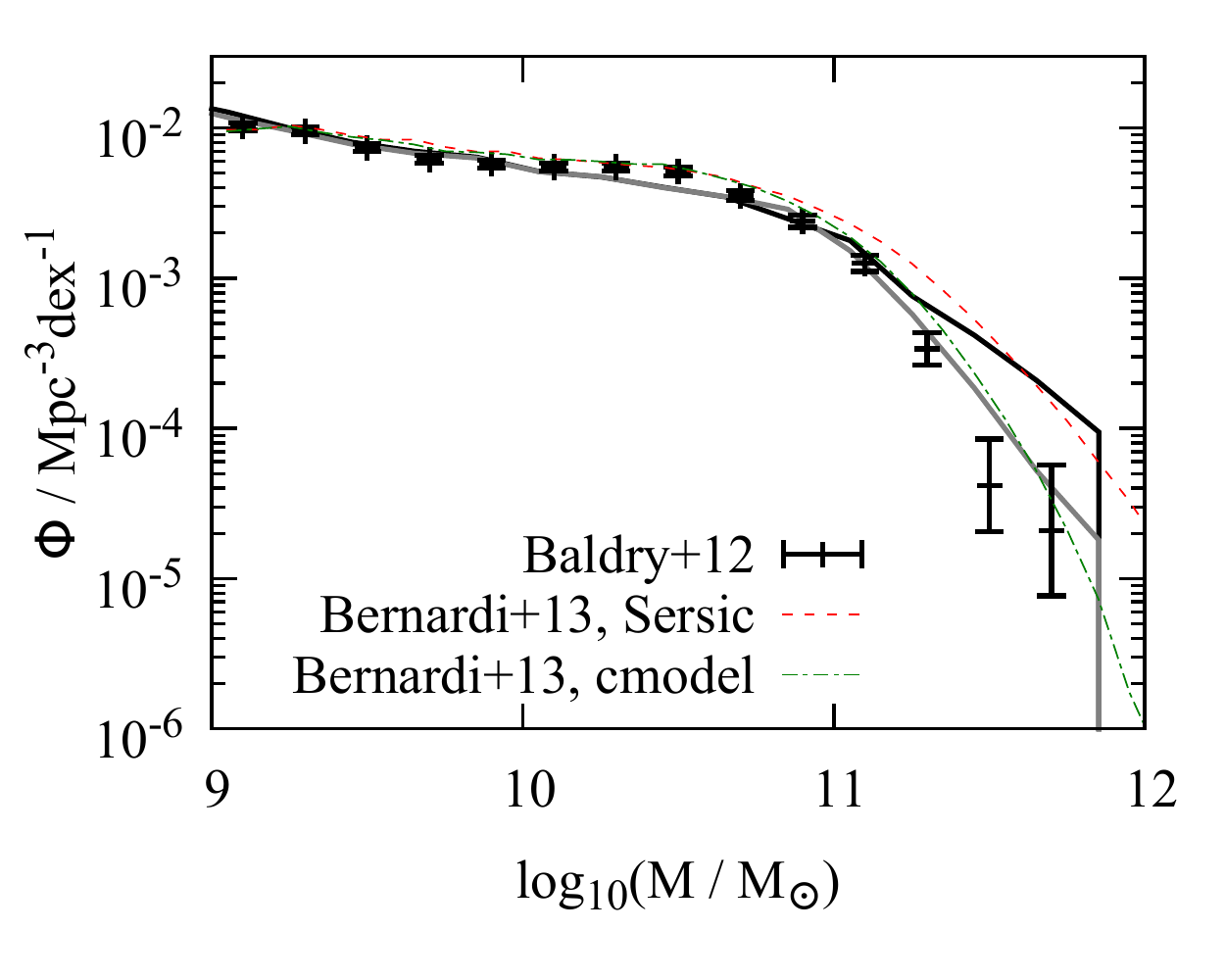}
\caption{The stellar mass function of galaxies predicted by 
Model III (grey line) compared to the prediction of the 
same model but assuming no stripping of satellites, i.e.
$f_{\rm TS}=1$ (the thick black line). The data points are the 
observational data of \citet{Baldry12}. The green dot-dashed line is 
the SMF obtained by \citet{Bernardi13} using SDSS `cmodel' 
magnitudes, while the red dashed line is the result obtained by the 
same authors using the magnitudes within the entire 
Sersic profiles (i.e. integrated to infinity) obtained by \citet{Simard11}.}
\label{hsmf}
\end{figure}

In our simple prescription for galaxy mergers, a constant 
fraction $f_{\rm TS}$ of the original stellar mass of the merging
satellite is accreted by the central, and the rest is deposited 
in a diffuse component, referred to as intracluster stars 
in clusters or as halo stars in general. $f_{\rm TS}$ is constrained to be 
about $40\%$, although the uncertainty is quite large (Tables 1 and 2).  
This low value of $f_{\rm TS}$ is driven predominantly 
by the observed SMFs at the massive end, i.e. with 
$M_\star \ga 10^{11}\Msun$ corresponding to $M_{\rm h}>10^{13}\Msunh$.
Had we set $f_{\rm TS}=1$, i.e. if all the original mass in an accreted 
satellite were added to the central, we would get a SMF that is  
significantly higher than the observations of \citet{Baldry12} 
at the very massive end (thick black line in Figure~\ref{hsmf}). 

Recent analyses show that the most massive galaxies 
(with $M_\star>10^{11}\Msun$) tend to have extended 
wings in their light profiles. This makes the luminosity 
(stellar mass) measurements of these galaxies quite uncertain. 
As shown by \citet{Bernardi13} and \citet{He13}, 
using different methods and light profiles to
fit galaxy images can lead to a factor of 2 difference
in the estimated luminosity of a massive galaxy. The
effect on the derived SMF can be seen in Figure~\ref{hsmf} 
comparing the red dashed line with the green dot-dashed line. 
It is interesting to note that our model prediction assuming $f_{\rm TS} = 1$
is consistent with the SMF obtained by \citet{Bernardi13} 
using the magnitudes within the entire Sersic profiles 
(the red line) obtained by \citet{Simard11}.
This suggests that the halo component defined in our model 
may simply be the extended profiles of massive galaxies
that are missing in the SMFs we use as constraints.   

\begin{figure}
\centering
\includegraphics[width=\linewidth]{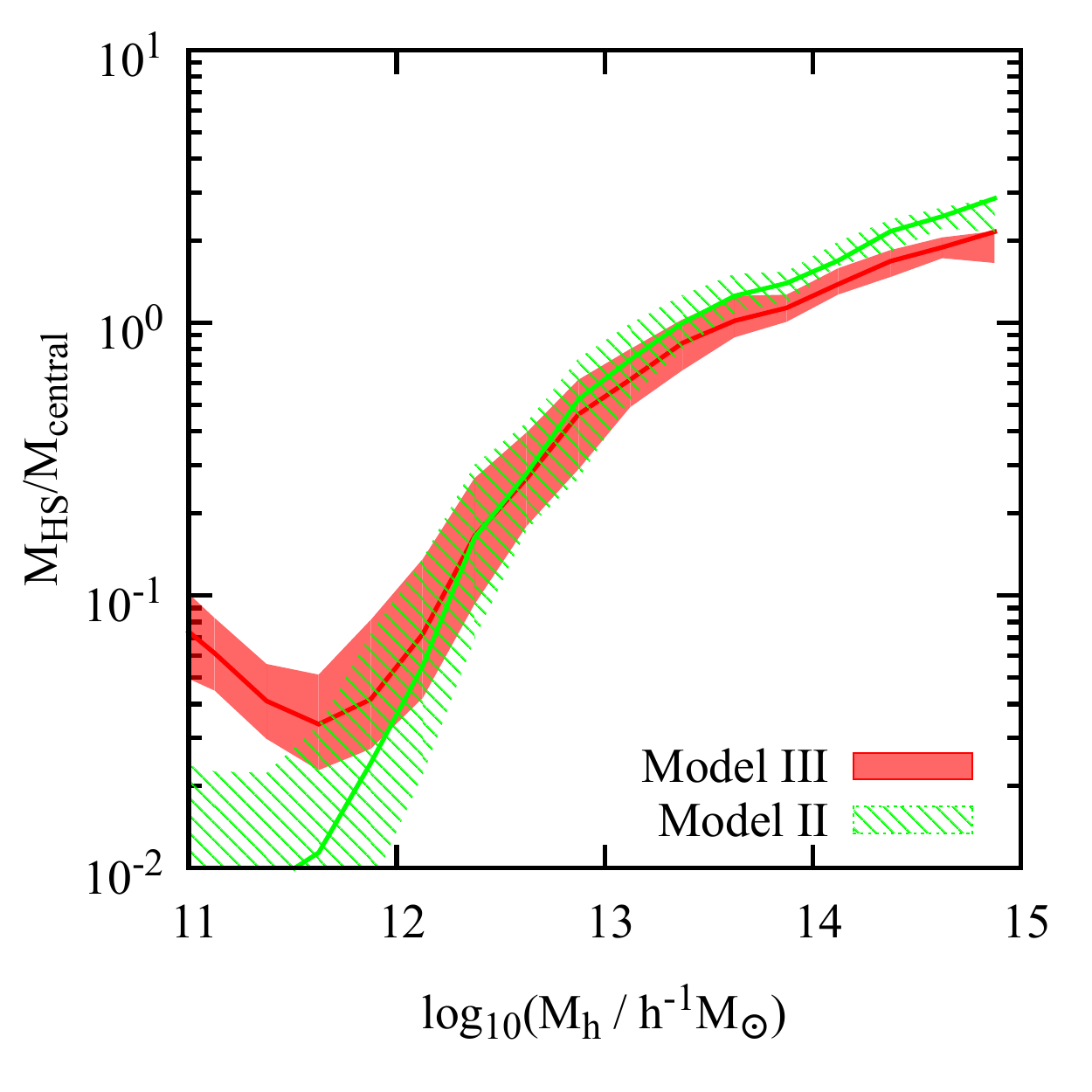}
\caption{The mass ratio between halo stars and the stellar mass 
of the central galaxy as a function of host halo mass at $z=0$. 
The green band is the prediction of Model II and the red band that
of Model III. Here again the widths of the bands represent the
variance among different halo merger trees.}
\label{f_hs}
\end{figure}

Figure\,\ref{f_hs} shows the ratio between the total mass of halo stars
and the stellar mass of the central galaxy as a function of host halo mass. 
In a cluster as massive as $10^{15}\Msunh$, the mass in the
diffuse component is about $2$ - $3$  times as high as that of the central
galaxy and in a group sized halo of mass $\sim 10^{13}\Msunh$, 
the ratio is about $1$. It drops rapidly towards lower 
halo mass. In a Milky Way mass halo, the predicted ratio is only a 
few percent. The flattening at the even lower mass end predicted 
by Model III owes to the boosted star formation at high $z$ in low-mass
haloes. 
The results for low-mass haloes should be taken with caution.  
As mentioned above, our model assumes a constant $f_{\rm TS}$ 
and it is constrained primarily by the SMFs at the massive end. 
It is unclear whether the same number also applies to galaxies with 
lower masses. 

\begin{figure}
\centering
\includegraphics[width=\linewidth]{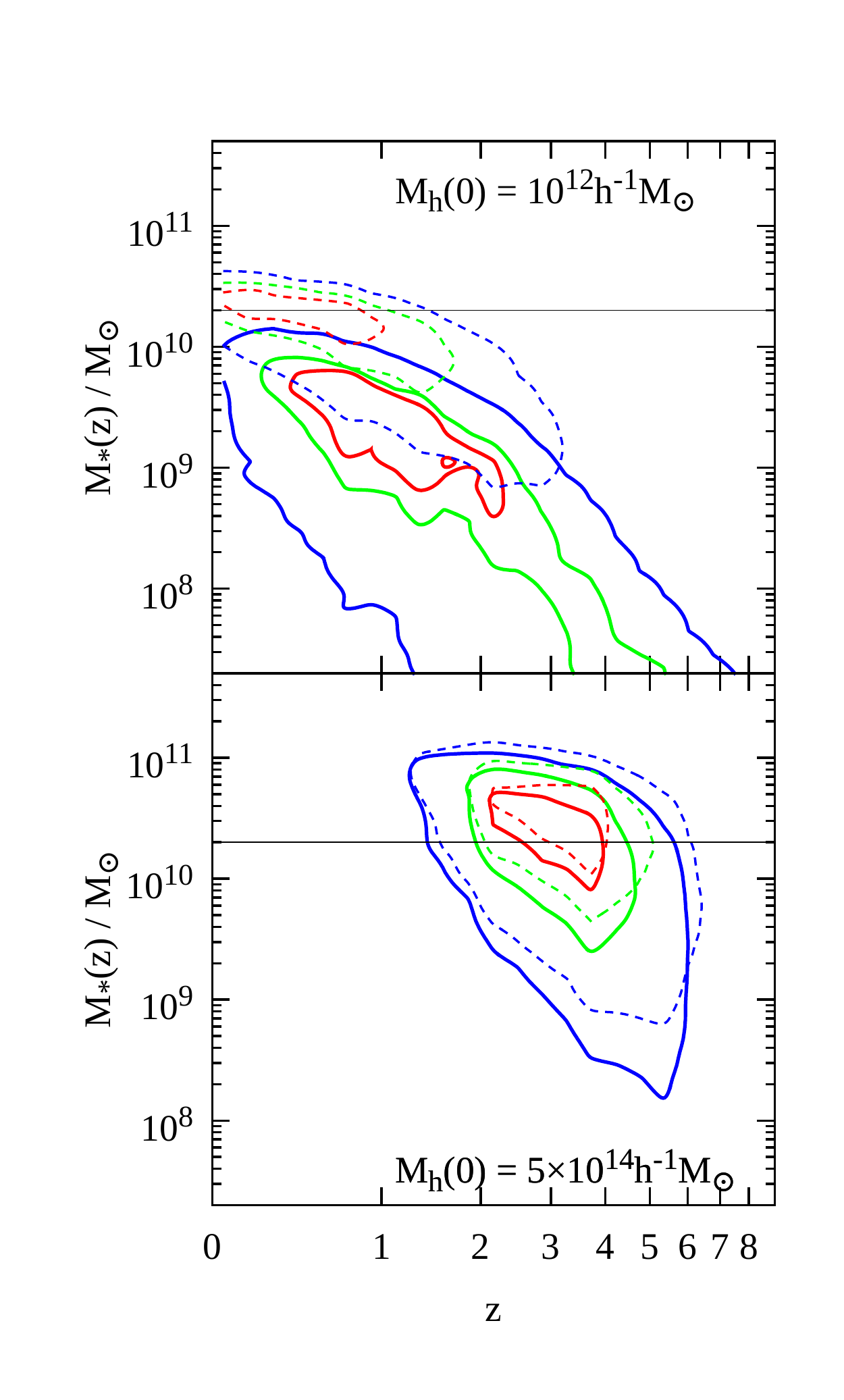}
\caption{The solid contours show the formation redshift  
and location (i.e. the stellar mass of the galaxy) 
of the halo stars. Red, green and
blue contours enclose 30\%, 60\% and 90\% of the
stars, respectively.  The dashed contours are the same 
as the solid ones but for stars in central galaxies.  
We only show Model III (Model II is similar). 
Results are shown for Milky Way mass haloes
(upper panel) and massive cluster haloes (lower panel). 
The horizontal lines mark $\Mstar=2\times 10^{10}\Msun$, 
above which the stellar metallicity begin to saturate  
\citep{Gallazzi05}.}
\label{hs}
\end{figure}

To understand the stellar population of the halo stars, 
we make a census of their formation time and location,   
in terms of redshift and the stellar mass 
of the host galaxy in which they formed, respectively. 
We show the result as the solid contours in Figure\,\ref{hs}.
For comparison, we also plot the results for stars in the 
central galaxies as the dashed contours. 
The stellar mass can be taken as crude proxy to stellar metallicity 
as the two are found to be correlated for local galaxies 
\citep{Gallazzi05}.  
The stellar metallicity increases with stellar mass for galaxies with 
stellar masses below $2\times10^{10}\Msun$ and becomes 
saturated at about solar metallicity above this mass 
(indicated by the horizontal lines in Figure\,\ref{hs}).

Compared to the stars in the central galaxy, the halo stars in a Milky Way 
sized halo form a distinct population: 
the mass weighted age of the halo stars is roughly about $9$-$10$ Gyr, 
in contrast with the central galaxy, which is about $6$ Gyr.
These halo stars formed in progenitors with mass lower than the horizontal line,
suggesting a metallicity much lower than the solar value. 
This is qualitatively consistent with the recent observations 
of M31 \citep{Bernard14}.
In massive clusters, on the other hand, 
the stellar populations in both the central galaxy
and the halo component are quite homogeneous, with ages of 
$\sim 10$ Gyr and with nearly solar metallicity. This prediction 
can be checked by studying the stellar age and metallicity of 
halo stars in clusters.

\section{Summary and Discussion}
\label{sec_summary}
In a previous paper \citep{Lu14}, we developed an empirical model 
to describe the star formation rates of central galaxies in haloes of 
different masses at different redshifts.  A series of nested models 
were constructed to accommodate more and more observational
constraints. We found that  Model II, which represents a class of 
`Slow Evolution' models in the literature,  can reproduce the 
SMFs since $z = 4$, but fails to accommodate the cluster galaxy 
luminosity function, the steep SFR-functions at high $z$, and 
the old stellar population seen in local dwarf galaxies. We also 
found that Model III is the simplest model family that can match 
all these observational data well. In the present paper, we use the same 
models, but with model parameters updated using recent observational
data of the  galaxy SMFs at high $z$. The results presented here 
confirm and re-enforce those of \citet{Lu14}. In particular, the constrained 
Model III predicts much steeper SMFs at $z\ge 4$ than Model II. 
Since the publication of \citet{Lu14}, two additional studies 
have presented evidence in support of this picture \citep{Madau14,Weisz14}.
These results suggest that the class of `Slow Evolution' models 
in the literature \citep{Yang13, Behroozi13a, Bethermin13}, in which  
star formation in haloes with $\Mhalo < 10^{11} \Msunh$ are 
suppressed at high $z$, are insufficient to describe the evolution 
of the galaxy population. 

We use our constrained model parameters to characterise 
the star formation and stellar mass assembly histories as well 
as the merger histories of galaxies of different masses. 
The results are summarised in the following.

First, the evolution of the galaxy population is found to be characterised by 
a number of characteristic halo mass scales:
\begin{enumerate}
\item 
A mass scale of $10^{13}\Msunh$ at $z=0$, decreasing  
to $\sim 3\times10^{12}\Msunh$ at $z>2$, above which {\it in situ} 
star formation drops rapidly, and the central galaxies experience 
frequent major mergers most of which will be dry.
\item 
A mass scale of $3\times10^{11}\Msunh$ at $z=0$, increasing  
to $10^{12}\Msunh$ at high $z$, at which the efficiency of {\it in situ} 
star formation reaches a maximum, with a SFR as high as about half 
of the baryon accretion rate into the host halo.  Major mergers are 
rare in this halo mass range.
\item
A mass scale of $10^{11}\Msunh$ at $z>3$, below which {\it in situ} star
formation has a rate about $0.1$ times the baryon accretion rate into
the host halo. On average, one or two major mergers are expected to 
occur for the central galaxies. 
\end{enumerate}

Second, galaxies hosted by haloes of different masses follow distinct
star formation and assembly histories.  Based on the characteristic
halo masses given above, central galaxies can be divided roughly into 
three different categories according to their formation and assembly
histories: 
\begin{enumerate}
\item 
For haloes with $M_{\rm h} > 10^{13}\Msunh$,  a strong {\it in situ} star
formation rate declines rapidly after reaching its peak value, and 
is followed by significant accretion of stars from satellites.  
For such massive systems, more massive galaxies tend to assemble 
their stellar mass later, contrary to the downsizing trends observed 
for lower mass galaxies.
\item 
For haloes with masses $10^{11}\Msunh < M_{\rm h} < 10^{13} \Msunh$,
mass assembly by accretion of satellites is not important, 
and the star formation is delayed relative to the formation of 
the host halo.
\item 
For haloes with masses below $10^{11}\Msunh$, assembly by 
accretion is again unimportant.  The star formation history is 
characterised by a burst at $z>2$ and a nearly constant star 
formation rate after $z=1$.  The relative importance of the 
early star formation increases with decreasing halo mass, 
and the `downsizing' trend is reversed.
\end{enumerate}

Third, we use the merger history of the model galaxies
to predict the bulge to total mass ratios of present galaxies. 
The average bulge mass fraction is found to depend strongly on halo mass:
\begin{enumerate}
\item 
In cluster sized haloes with $M_{\rm h}>3\times10^{13}\Msunh$, 
almost all the centrals are ellipticals formed through frequent major mergers.
\item 
In group sized haloes with masses between $3\times10^{12}\Msunh$
and $3\times10^{13}\Msunh$, the distribution of the bulge-to-total 
ratio of the central galaxies is strongly bimodal. 
Those galaxies that experienced a recent major merger are
spheroid dominated, whereas the others are free of any significant
merger-driven bulge.
\item 
For haloes with masses $3\times10^{11}
\Msunh<M_{\rm h}<3\times10^{12}\Msunh$,  
central galaxies with a significant merger-driven bulge are extremely rare.
\item 
For dwarf galaxies, half of them have significant 
(with $B/T>10\%$) spheroidal components formed during their 
early star burst phase ($z>z_{c}\approx2$). Satellite galaxies of similar masses 
tend to have a larger bulge fraction than centrals. 
\end{enumerate}
We emphasise again that bulges can form in various other ways 
than through major mergers, and that our prediction only applies 
to major merger-driven bulges. 

Finally, we have made predictions for the amount of halo stars, and when 
and where these stars form in comparison with stars in the 
corresponding central galaxies. The results are: 
\begin{enumerate}
\item 
In a Milky Way mass halo, the total mass in halo stars is 
2 to 5 percent of the mass of the central galaxy, and this number 
increases to $\sim 100\%$ in haloes with $M_{\rm h}\sim 10^{13}\Msunh$,
and to about $200\%$ to $300\%$ in massive clusters.
\item 
In a Milky Way mass halo, the stars in the central galaxy and halo
stars form two distinct stellar populations, with the latter being older 
and poorer in metals. 
In contrast, these two components form a quite homogeneous 
population (old and with solar metallicity) in massive clusters.
\end{enumerate}

All these results are obtained in an empirical way, independent of 
any detailed assumptions about the underlying physical processes 
that drive the evolution of the galaxy population. Clearly, our
results should be compared with the predictions of numerical 
simulations and/or semi-analytical models to 
constrain theories of galaxy formation. We will come back to
this in a forthcoming paper.   
   
\section*{Acknowledgements}
We thank Stephane Charlot for discussion, and Ryan Quadri for
providing the ZFOURGE/CANDELS stellar mass functions. 
We would like to acknowledge the support of NSF AST-1109354 and NSF
AST-0908334.


\label{lastpage}
\end{document}

%% file: macro.tex
\def\beq{\begin{equation}}
\def\eeq{\end{equation}}
\def\bey{\begin{eqnarray}}
\def\eey{\end{eqnarray}}

\def\Msun{\,{\rm M_\odot}}
\def\Msunh{\, h^{-1}{\rm M_\odot}}
\def\Msunhh{\, h^{-2}{\rm M_\odot}}
\def\Mhalo{\, M_{\rm h}}
\def\Mstar{\, M_\star}
\def\zhalf{\, z_{0.5}}
\def\zten{\, z_{0.1}}

\def\gs{\mathrel{\raise1.16pt\hbox{$>$}\kern-7.0pt
\lower3.06pt\hbox{{$\scriptstyle \sim$}}}}
\def\ls{\mathrel{\raise1.16pt\hbox{$<$}\kern-7.0pt
\lower3.06pt\hbox{{$\scriptstyle \sim$}}}}
\def\gtsima{$\; \buildrel > \over \sim \;$}
\def\ltsima{$\; \buildrel < \over \sim \;$}
\def\prosima{$\; \buildrel \propto \over \sim \;$}
\def\gsim{\lower.5ex\hbox{\gtsima}}
\def\lsim{\lower.5ex\hbox{\ltsima}}
\def\simgt{\lower.5ex\hbox{\gtsima}}
\def\simlt{\lower.5ex\hbox{\ltsima}}
\def\simpr{\lower.5ex\hbox{\prosima}}
\def\la{\lsim}
\def\ga{\gsim}
